\documentclass[longauth,pdfencoding=auto,psdextra]{aa}

\usepackage{graphicx}
\usepackage{natbib}
\usepackage{scalerel}

\usepackage[table]{xcolor}
\usepackage{comment}
\usepackage{orcidlink} 	

\usepackage{txfonts}

\hypersetup{
    colorlinks=true,
    linkcolor=blue,
    filecolor=magenta,      
    urlcolor=blue,
    citecolor=blue
}
\makeatletter
\renewcommand*\aa@pageof{, page \thepage{} of \pageref*{LastPage}}
\makeatother

\usepackage[utf8]{inputenc}
\usepackage[switch, modulo]{lineno}
\renewcommand{\linenumbers}[0]{}

\usepackage{euclid}

\usepackage[nameinlink,capitalise]{cleveref}

\crefname{section}{Sect.}{Sects.}
\Crefname{section}{Section}{Sections}
\crefname{figure}{Fig.}{Figs.}
\Crefname{figure}{Figure}{Figures}
\crefname{equation}{Eq.}{Eqs.}
\Crefname{equation}{Equation}{Equations}
\crefname{table}{Table}{Tables}
\crefname{appendix}{Appendix}{Appendices}

\defcitealias{Q1-TP003}{NIRQ1}

\usepackage[nolist,nohyperlinks]{acronym}

\begin{document}

\acrodef{BPM}{bad-pixel mask}
\acrodef{CR}{cosmic ray}
\acrodef{DP}{data product}
\acrodef{DQ}{data quality}
\acrodef{EAS}{Euclid Archive System}
\acrodef{EDF}{Euclid Deep Field}
\acrodef{EDS}{Euclid Deep Survey}
\acrodef{EWS}{Euclid Wide Survey}
\acrodef{FPA}{focal-plane array}
\acrodef{HDU}{header data unit}
\acrodef{HST}{\textit{Hubble} Space Telescope}
\acrodef{IS}{inverse sensitivity}
\acrodef{JWST}{\textit {James Webb} Space Telescope}
\acrodef{LED}{light-emitting diode}
\acrodef{MAD}{median absolute deviation}
\acrodef{MEF}{multi-extension FITS}
\acrodef{NIR}{near-infrared}
\acrodef{NIR PF}{near-infrared processing function}
\acrodef{NISP}{Near-Infrared Spectrometer and Photometer}
\acrodef{PE}{processing element}
\acrodef{PSF}{point-spread function}
\acrodef{PV}{performance verification}
\acrodef{QE}{quantum efficiency}
\acrodef{ROS}{reference observing sequence}
\acrodef{SED}{spectral energy distribution}
\acrodef{SGS}{Science Ground Segment}
\acrodef{WD}{white dwarf}

\title{Euclid Data Release 1 (DR1)}
\subtitle{Near-infrared processing and data products}  

\newcommand{\orcid}[1]{} 

\author{Euclid Collaboration: G.~Polenta\orcid{0000-0003-4067-9196}\thanks{\email{gianluca.polenta@asi.it}}\inst{\ref{aff1}}
\and M.~Frailis\orcid{0000-0002-7400-2135}\inst{\ref{aff2}}
\and A.~Alavi\orcid{0000-0002-8630-6435}\inst{\ref{aff3}}
\and P.~N.~Appleton\orcid{0000-0002-7607-8766}\inst{\ref{aff3}}
\and P.~Awad\orcid{0000-0002-0428-849X}\inst{\ref{aff4}}
\and A.~Bonchi\orcid{0000-0002-2667-5482}\inst{\ref{aff1}}
\and L.~Bramante\inst{\ref{aff5}}
\and G.~Calderone\orcid{0000-0002-7738-5389}\inst{\ref{aff2}}
\and F.~Cogato\orcid{0000-0003-4632-6113}\inst{\ref{aff6}}
\and S.~Conseil\orcid{0000-0002-3657-4191}\inst{\ref{aff7},\ref{aff8}}
\and M.~Correnti\orcid{0000-0001-6464-3257}\inst{\ref{aff9},\ref{aff1}}
\and R.~da~Silva\orcid{0000-0003-4788-677X}\inst{\ref{aff9},\ref{aff1}}
\and I.~Das\orcid{0009-0007-7088-2044}\inst{\ref{aff3}}
\and F.~Faustini\orcid{0000-0001-6274-5145}\inst{\ref{aff9}}
\and Y.~Fu\orcid{0000-0002-0759-0504}\inst{\ref{aff4},\ref{aff10}}
\and T.~Gasparetto\orcid{0000-0002-7913-4866}\inst{\ref{aff9}}
\and W.~Gillard\orcid{0000-0003-4744-9748}\inst{\ref{aff11}}
\and S.~Hemmati\orcid{0000-0003-2226-5395}\inst{\ref{aff3}}
\and J.~Jacobson\inst{\ref{aff3}}
\and K.~Jahnke\orcid{0000-0003-3804-2137}\inst{\ref{aff12}}
\and B.~Kubik\orcid{0009-0006-5823-4880}\inst{\ref{aff8}}
\and X.~Liu\inst{\ref{aff3}}
\and C.~Macabiau\inst{\ref{aff8}}
\and E.~Medinaceli\orcid{0000-0002-4040-7783}\inst{\ref{aff6}}
\and P.~W.~Morris\orcid{0000-0002-5186-4381}\inst{\ref{aff13}}
\and K.~Paterson\orcid{0000-0001-8340-3486}\inst{\ref{aff12}}
\and M.~Radovich\orcid{0000-0002-3585-866X}\inst{\ref{aff14}}
\and M.~Schirmer\orcid{0000-0003-2568-9994}\inst{\ref{aff12}}
\and H.~I.~Teplitz\orcid{0000-0002-7064-5424}\inst{\ref{aff15}}
\and A.~Troja\orcid{0000-0003-0239-4595}\inst{\ref{aff2}}
\and B.~P.~Venemans\orcid{0000-0001-9024-8322}\inst{\ref{aff4}}
\and B.~Altieri\orcid{0000-0003-3936-0284}\inst{\ref{aff16}}
\and S.~Andreon\orcid{0000-0002-2041-8784}\inst{\ref{aff17}}
\and N.~Auricchio\orcid{0000-0003-4444-8651}\inst{\ref{aff6}}
\and H.~Aussel\orcid{0000-0002-1371-5705}\inst{\ref{aff18}}
\and C.~Baccigalupi\orcid{0000-0002-8211-1630}\inst{\ref{aff19},\ref{aff2},\ref{aff20},\ref{aff21}}
\and M.~Baldi\orcid{0000-0003-4145-1943}\inst{\ref{aff22},\ref{aff6},\ref{aff23}}
\and A.~Balestra\orcid{0000-0002-6967-261X}\inst{\ref{aff14}}
\and S.~Bardelli\orcid{0000-0002-8900-0298}\inst{\ref{aff6}}
\and P.~Battaglia\orcid{0000-0002-7337-5909}\inst{\ref{aff6}}
\and A.~N.~Belikov\inst{\ref{aff10},\ref{aff24}}
\and R.~Bender\orcid{0000-0001-7179-0626}\inst{\ref{aff25},\ref{aff26}}
\and A.~Biviano\orcid{0000-0002-0857-0732}\inst{\ref{aff2},\ref{aff19}}
\and E.~Branchini\orcid{0000-0002-0808-6908}\inst{\ref{aff27},\ref{aff28},\ref{aff17}}
\and M.~Brescia\orcid{0000-0001-9506-5680}\inst{\ref{aff29},\ref{aff30}}
\and J.~Brinchmann\orcid{0000-0003-4359-8797}\inst{\ref{aff31},\ref{aff32},\ref{aff33}}
\and S.~Camera\orcid{0000-0003-3399-3574}\inst{\ref{aff34},\ref{aff35},\ref{aff36}}
\and G.~Ca\~nas-Herrera\orcid{0000-0003-2796-2149}\inst{\ref{aff4}}
\and V.~Capobianco\orcid{0000-0002-3309-7692}\inst{\ref{aff36}}
\and C.~Carbone\orcid{0000-0003-0125-3563}\inst{\ref{aff37}}
\and V.~F.~Cardone\inst{\ref{aff9},\ref{aff38}}
\and J.~Carretero\orcid{0000-0002-3130-0204}\inst{\ref{aff39},\ref{aff40}}
\and F.~J.~Castander\orcid{0000-0001-7316-4573}\inst{\ref{aff41},\ref{aff42}}
\and M.~Castellano\orcid{0000-0001-9875-8263}\inst{\ref{aff9}}
\and G.~Castignani\orcid{0000-0001-6831-0687}\inst{\ref{aff6}}
\and S.~Cavuoti\orcid{0000-0002-3787-4196}\inst{\ref{aff30},\ref{aff43}}
\and K.~C.~Chambers\orcid{0000-0001-6965-7789}\inst{\ref{aff44}}
\and A.~Cimatti\inst{\ref{aff45}}
\and C.~Colodro-Conde\inst{\ref{aff46}}
\and G.~Congedo\orcid{0000-0003-2508-0046}\inst{\ref{aff47}}
\and C.~J.~Conselice\orcid{0000-0003-1949-7638}\inst{\ref{aff48}}
\and L.~Conversi\orcid{0000-0002-6710-8476}\inst{\ref{aff49},\ref{aff16}}
\and Y.~Copin\orcid{0000-0002-5317-7518}\inst{\ref{aff8}}
\and F.~Courbin\orcid{0000-0003-0758-6510}\inst{\ref{aff50},\ref{aff51},\ref{aff52}}
\and H.~M.~Courtois\orcid{0000-0003-0509-1776}\inst{\ref{aff53}}
\and M.~Cropper\orcid{0000-0003-4571-9468}\inst{\ref{aff54}}
\and A.~Da~Silva\orcid{0000-0002-6385-1609}\inst{\ref{aff55},\ref{aff56}}
\and H.~Degaudenzi\orcid{0000-0002-5887-6799}\inst{\ref{aff57}}
\and G.~De~Lucia\orcid{0000-0002-6220-9104}\inst{\ref{aff2}}
\and H.~Dole\orcid{0000-0002-9767-3839}\inst{\ref{aff58},\ref{aff59}}
\and F.~Dubath\orcid{0000-0002-6533-2810}\inst{\ref{aff57}}
\and X.~Dupac\inst{\ref{aff16}}
\and S.~Dusini\orcid{0000-0002-1128-0664}\inst{\ref{aff60}}
\and S.~Escoffier\orcid{0000-0002-2847-7498}\inst{\ref{aff11}}
\and M.~Fabricius\orcid{0000-0002-7025-6058}\inst{\ref{aff25},\ref{aff26}}
\and M.~Farina\orcid{0000-0002-3089-7846}\inst{\ref{aff61}}
\and R.~Farinelli\inst{\ref{aff6}}
\and S.~Ferriol\inst{\ref{aff8}}
\and F.~Finelli\orcid{0000-0002-6694-3269}\inst{\ref{aff6},\ref{aff62}}
\and S.~Fotopoulou\orcid{0000-0002-9686-254X}\inst{\ref{aff63}}
\and N.~Fourmanoit\orcid{0009-0005-6816-6925}\inst{\ref{aff11}}
\and E.~Franceschi\orcid{0000-0002-0585-6591}\inst{\ref{aff6}}
\and M.~Fumana\orcid{0000-0001-6787-5950}\inst{\ref{aff37}}
\and L.~Gabarra\orcid{0000-0002-8486-8856}\inst{\ref{aff64}}
\and S.~Galeotta\orcid{0000-0002-3748-5115}\inst{\ref{aff2}}
\and K.~George\orcid{0000-0002-1734-8455}\inst{\ref{aff65}}
\and C.~Giocoli\orcid{0000-0002-9590-7961}\inst{\ref{aff6},\ref{aff23}}
\and J.~Gracia-Carpio\orcid{0000-0003-4689-3134}\inst{\ref{aff25}}
\and B.~R.~Granett\orcid{0000-0003-2694-9284}\inst{\ref{aff17}}
\and A.~Grazian\orcid{0000-0002-5688-0663}\inst{\ref{aff14}}
\and F.~Grupp\inst{\ref{aff25},\ref{aff26}}
\and L.~Guzzo\orcid{0000-0001-8264-5192}\inst{\ref{aff66},\ref{aff17},\ref{aff67}}
\and W.~G.~Hartley\orcid{0000-0002-4563-5101}\inst{\ref{aff57}}
\and S.~V.~H.~Haugan\orcid{0000-0001-9648-7260}\inst{\ref{aff68}}
\and J.~Hoar\inst{\ref{aff16}}
\and H.~Hoekstra\orcid{0000-0002-0641-3231}\inst{\ref{aff4}}
\and W.~Holmes\orcid{0009-0007-8554-4646}\inst{\ref{aff69}}
\and I.~M.~Hook\orcid{0000-0002-2960-978X}\inst{\ref{aff70}}
\and F.~Hormuth\inst{\ref{aff71}}
\and A.~Hornstrup\orcid{0000-0002-3363-0936}\inst{\ref{aff72},\ref{aff73}}
\and M.~Huertas-Company\orcid{0000-0002-1416-8483}\inst{\ref{aff46},\ref{aff74},\ref{aff75}}
\and M.~Jhabvala\inst{\ref{aff76}}
\and S.~Kermiche\orcid{0000-0002-0302-5735}\inst{\ref{aff11}}
\and A.~Kiessling\orcid{0000-0002-2590-1273}\inst{\ref{aff69}}
\and R.~Kohley\inst{\ref{aff16}}
\and M.~K\"ummel\orcid{0000-0003-2791-2117}\inst{\ref{aff26}}
\and M.~Kunz\orcid{0000-0002-3052-7394}\inst{\ref{aff77}}
\and H.~Kurki-Suonio\orcid{0000-0002-4618-3063}\inst{\ref{aff78},\ref{aff79}}
\and A.~M.~C.~Le~Brun\orcid{0000-0002-0936-4594}\inst{\ref{aff80}}
\and V.~Le~Brun\orcid{0000-0002-5027-1939}\inst{\ref{aff81}}
\and D.~Le~Mignant\orcid{0000-0002-5339-5515}\inst{\ref{aff81}}
\and P.~Liebing\orcid{0009-0005-8043-7150}\inst{\ref{aff54}}
\and S.~Ligori\orcid{0000-0003-4172-4606}\inst{\ref{aff36}}
\and P.~B.~Lilje\orcid{0000-0003-4324-7794}\inst{\ref{aff68}}
\and V.~Lindholm\orcid{0000-0003-2317-5471}\inst{\ref{aff78},\ref{aff79}}
\and I.~Lloro\orcid{0000-0001-5966-1434}\inst{\ref{aff82}}
\and M.~Magliocchetti\orcid{0000-0001-9158-4838}\inst{\ref{aff61}}
\and G.~Mainetti\orcid{0000-0003-2384-2377}\inst{\ref{aff83}}
\and O.~Mansutti\orcid{0000-0001-5758-4658}\inst{\ref{aff2}}
\and S.~Marcin\inst{\ref{aff84}}
\and O.~Marggraf\orcid{0000-0001-7242-3852}\inst{\ref{aff85}}
\and M.~Martinelli\orcid{0000-0002-6943-7732}\inst{\ref{aff9},\ref{aff38}}
\and N.~Martinet\orcid{0000-0003-2786-7790}\inst{\ref{aff81}}
\and F.~Marulli\orcid{0000-0002-8850-0303}\inst{\ref{aff86},\ref{aff6},\ref{aff23}}
\and R.~J.~Massey\orcid{0000-0002-6085-3780}\inst{\ref{aff87}}
\and N.~Mauri\orcid{0000-0001-8196-1548}\inst{\ref{aff45},\ref{aff23}}
\and H.~J.~McCracken\orcid{0000-0002-9489-7765}\inst{\ref{aff88}}
\and C.~J.~R.~McPartland\orcid{0000-0003-0639-025X}\inst{\ref{aff73},\ref{aff89}}
\and S.~Mei\orcid{0000-0002-2849-559X}\inst{\ref{aff90},\ref{aff91}}
\and M.~Meneghetti\orcid{0000-0003-1225-7084}\inst{\ref{aff6},\ref{aff23}}
\and E.~Merlin\orcid{0000-0001-6870-8900}\inst{\ref{aff14}}
\and G.~Meylan\orcid{0000-0001-6503-0209}\inst{\ref{aff92}}
\and P.~Monaco\orcid{0000-0003-2083-7564}\inst{\ref{aff93},\ref{aff2},\ref{aff20},\ref{aff19}}
\and A.~Mora\orcid{0000-0002-1922-8529}\inst{\ref{aff94}}
\and M.~Moresco\orcid{0000-0002-7616-7136}\inst{\ref{aff86},\ref{aff6}}
\and C.~Moretti\orcid{0000-0003-3314-8936}\inst{\ref{aff2},\ref{aff19},\ref{aff20}}
\and L.~Moscardini\orcid{0000-0002-3473-6716}\inst{\ref{aff86},\ref{aff6},\ref{aff23}}
\and R.~Nakajima\orcid{0009-0009-1213-7040}\inst{\ref{aff85}}
\and C.~Neissner\orcid{0000-0001-8524-4968}\inst{\ref{aff95},\ref{aff40}}
\and R.~C.~Nichol\orcid{0000-0003-0939-6518}\inst{\ref{aff96}}
\and S.-M.~Niemi\orcid{0009-0005-0247-0086}\inst{\ref{aff97}}
\and C.~Padilla\orcid{0000-0001-7951-0166}\inst{\ref{aff95}}
\and S.~Paltani\orcid{0000-0002-8108-9179}\inst{\ref{aff57}}
\and F.~Pasian\orcid{0000-0002-4869-3227}\inst{\ref{aff2}}
\and J.~A.~Peacock\orcid{0000-0002-1168-8299}\inst{\ref{aff47}}
\and W.~J.~Percival\orcid{0000-0002-0644-5727}\inst{\ref{aff98},\ref{aff99},\ref{aff100}}
\and V.~Pettorino\orcid{0000-0002-4203-9320}\inst{\ref{aff97}}
\and S.~Pires\orcid{0000-0002-0249-2104}\inst{\ref{aff18}}
\and M.~Poncet\inst{\ref{aff101}}
\and L.~A.~Popa\inst{\ref{aff102}}
\and F.~Raison\orcid{0000-0002-7819-6918}\inst{\ref{aff25}}
\and A.~Refregier\inst{\ref{aff103}}
\and A.~Renzi\orcid{0000-0001-9856-1970}\inst{\ref{aff104},\ref{aff60},\ref{aff6}}
\and J.~Rhodes\orcid{0000-0002-4485-8549}\inst{\ref{aff69}}
\and G.~Riccio\inst{\ref{aff30}}
\and I.~Risso\orcid{0000-0003-2525-7761}\inst{\ref{aff27},\ref{aff28},\ref{aff17}}
\and F.~Rizzo\orcid{0000-0002-9407-585X}\inst{\ref{aff2}}
\and E.~Romelli\orcid{0000-0003-3069-9222}\inst{\ref{aff2}}
\and C.~Rosset\orcid{0000-0003-0286-2192}\inst{\ref{aff90}}
\and B.~Rusholme\orcid{0000-0001-7648-4142}\inst{\ref{aff3}}
\and R.~Saglia\orcid{0000-0003-0378-7032}\inst{\ref{aff26},\ref{aff25}}
\and Z.~Sakr\orcid{0000-0002-4823-3757}\inst{\ref{aff105},\ref{aff106},\ref{aff107}}
\and A.~G.~S\'anchez\orcid{0000-0003-1198-831X}\inst{\ref{aff25}}
\and D.~Sapone\orcid{0000-0001-7089-4503}\inst{\ref{aff108}}
\and B.~Sartoris\orcid{0000-0003-1337-5269}\inst{\ref{aff109}}
\and P.~Schneider\orcid{0000-0001-8561-2679}\inst{\ref{aff85}}
\and T.~Schrabback\orcid{0000-0002-6987-7834}\inst{\ref{aff110}}
\and M.~Scodeggio\inst{\ref{aff37}}
\and A.~Secroun\orcid{0000-0003-0505-3710}\inst{\ref{aff11}}
\and E.~Sefusatti\orcid{0000-0003-0473-1567}\inst{\ref{aff2},\ref{aff19},\ref{aff20}}
\and E.~Sihvola\orcid{0000-0003-1804-7715}\inst{\ref{aff78},\ref{aff79}}
\and P.~Simon\inst{\ref{aff85}}
\and C.~Sirignano\orcid{0000-0002-0995-7146}\inst{\ref{aff104},\ref{aff60}}
\and G.~Sirri\orcid{0000-0003-2626-2853}\inst{\ref{aff23}}
\and L.~Stanco\orcid{0000-0002-9706-5104}\inst{\ref{aff60}}
\and P.~Tallada-Cresp\'{i}\orcid{0000-0002-1336-8328}\inst{\ref{aff39},\ref{aff40}}
\and A.~N.~Taylor\inst{\ref{aff47}}
\and I.~Tereno\orcid{0000-0002-4537-6218}\inst{\ref{aff55},\ref{aff56}}
\and N.~Tessore\orcid{0000-0002-9696-7931}\inst{\ref{aff54}}
\and S.~Toft\orcid{0000-0003-3631-7176}\inst{\ref{aff111},\ref{aff89}}
\and R.~Toledo-Moreo\orcid{0000-0002-2997-4859}\inst{\ref{aff112},\ref{aff113}}
\and F.~Torradeflot\orcid{0000-0003-1160-1517}\inst{\ref{aff40},\ref{aff39}}
\and A.~Tsyganov\inst{\ref{aff114}}
\and I.~Tutusaus\orcid{0000-0002-3199-0399}\inst{\ref{aff41},\ref{aff42},\ref{aff106}}
\and J.~Valiviita\orcid{0000-0001-6225-3693}\inst{\ref{aff78},\ref{aff79}}
\and T.~Vassallo\orcid{0000-0001-6512-6358}\inst{\ref{aff2},\ref{aff65}}
\and A.~Veropalumbo\orcid{0000-0003-2387-1194}\inst{\ref{aff17},\ref{aff28},\ref{aff27}}
\and D.~Vibert\orcid{0009-0008-0607-631X}\inst{\ref{aff81}}
\and Y.~Wang\orcid{0000-0002-4749-2984}\inst{\ref{aff3}}
\and J.~Weller\orcid{0000-0002-8282-2010}\inst{\ref{aff26},\ref{aff25}}
\and A.~Zacchei\orcid{0000-0003-0396-1192}\inst{\ref{aff2},\ref{aff19}}
\and G.~Zamorani\orcid{0000-0002-2318-301X}\inst{\ref{aff6}}
\and F.~M.~Zerbi\orcid{0000-0002-9996-973X}\inst{\ref{aff17}}
\and E.~Zucca\orcid{0000-0002-5845-8132}\inst{\ref{aff6}}
\and I.~T.~Andika\orcid{0000-0001-6102-9526}\inst{\ref{aff26}}
\and M.~Ballardini\orcid{0000-0003-4481-3559}\inst{\ref{aff115},\ref{aff116},\ref{aff6}}
\and P.~Bergamini\orcid{0000-0003-1383-9414}\inst{\ref{aff6}}
\and S.~Borgani\orcid{0000-0001-6151-6439}\inst{\ref{aff93},\ref{aff19},\ref{aff2},\ref{aff20},\ref{aff117}}
\and E.~Bozzo\orcid{0000-0002-8201-1525}\inst{\ref{aff57}}
\and C.~Burigana\orcid{0000-0002-3005-5796}\inst{\ref{aff118},\ref{aff62}}
\and R.~Cabanac\orcid{0000-0001-6679-2600}\inst{\ref{aff106}}
\and M.~Calabrese\orcid{0000-0002-2637-2422}\inst{\ref{aff119},\ref{aff37}}
\and A.~Cappi\inst{\ref{aff109},\ref{aff6}}
\and F.~Caro\orcid{0009-0003-1053-0507}\inst{\ref{aff9}}
\and T.~Castro\orcid{0000-0002-6292-3228}\inst{\ref{aff120},\ref{aff2},\ref{aff19}}
\and D.~Di~Ferdinando\inst{\ref{aff23}}
\and J.~A.~Escartin~Vigo\inst{\ref{aff25}}
\and G.~Fabbian\orcid{0000-0002-3255-4695}\inst{\ref{aff58}}
\and J.~Garc\'ia-Bellido\orcid{0000-0002-9370-8360}\inst{\ref{aff105}}
\and V.~Gautard\inst{\ref{aff121}}
\and A.~Gregorio\orcid{0000-0003-4028-8785}\inst{\ref{aff93},\ref{aff2},\ref{aff20}}
\and C.~M.~Gutierrez\orcid{0000-0001-7854-783X}\inst{\ref{aff46},\ref{aff122}}
\and A.~Hall\orcid{0000-0002-3139-8651}\inst{\ref{aff47}}
\and Y.~Kang\orcid{0009-0000-8588-7250}\inst{\ref{aff110}}
\and G.~Leroy\orcid{0009-0004-2523-4425}\inst{\ref{aff123},\ref{aff87}}
\and A.~Loureiro\orcid{0000-0002-4371-0876}\inst{\ref{aff124},\ref{aff125}}
\and J.~Macias-Perez\orcid{0000-0002-5385-2763}\inst{\ref{aff126}}
\and C.~Mancini\orcid{0000-0002-4297-0561}\inst{\ref{aff37}}
\and R.~Maoli\orcid{0000-0002-6065-3025}\inst{\ref{aff127},\ref{aff9}}
\and R.~B.~Metcalf\orcid{0000-0003-3167-2574}\inst{\ref{aff86},\ref{aff6}}
\and S.~Nadathur\orcid{0000-0001-9070-3102}\inst{\ref{aff128}}
\and L.~Patrizii\inst{\ref{aff23}}
\and M.~P\"ontinen\orcid{0000-0001-5442-2530}\inst{\ref{aff78}}
\and E.~Sarpa\orcid{0000-0002-1256-655X}\inst{\ref{aff2}}
\and V.~Scottez\orcid{0009-0008-3864-940X}\inst{\ref{aff129},\ref{aff130}}
\and M.~Sereno\orcid{0000-0003-0302-0325}\inst{\ref{aff6},\ref{aff23}}
\and K.~Tanidis\orcid{0000-0001-9843-5130}\inst{\ref{aff131}}
\and M.~Tenti\orcid{0000-0002-4254-5901}\inst{\ref{aff23}}
\and M.~Tucci\inst{\ref{aff57}}
\and M.~Viel\orcid{0000-0002-2642-5707}\inst{\ref{aff19},\ref{aff2},\ref{aff21},\ref{aff20},\ref{aff117}}
\and M.~Wiesmann\orcid{0009-0000-8199-5860}\inst{\ref{aff68}}
\and Abdurro'uf\orcid{0000-0002-5258-8761}\inst{\ref{aff132}}
\and J.~A.~Acevedo~Barroso\orcid{0000-0002-9654-1711}\inst{\ref{aff69}}
\and Y.~Akrami\orcid{0000-0002-2407-7956}\inst{\ref{aff105},\ref{aff133}}
\and G.~Alguero\inst{\ref{aff126}}
\and G.~Angora\orcid{0000-0002-0316-6562}\inst{\ref{aff30},\ref{aff115}}
\and S.~Anselmi\orcid{0000-0002-3579-9583}\inst{\ref{aff60},\ref{aff104},\ref{aff134}}
\and M.~Archidiacono\orcid{0000-0003-4952-9012}\inst{\ref{aff66},\ref{aff67}}
\and G.~Aric\`o\orcid{0000-0002-2802-2928}\inst{\ref{aff23}}
\and F.~Atrio-Barandela\orcid{0000-0002-2130-2513}\inst{\ref{aff135}}
\and M.~Aubert\orcid{0009-0002-7667-8814}\inst{\ref{aff8}}
\and M.~Baes\orcid{0000-0002-3930-2757}\inst{\ref{aff136}}
\and N.~Bartolo\inst{\ref{aff104},\ref{aff60},\ref{aff14}}
\and L.~Bazzanini\orcid{0000-0003-0727-0137}\inst{\ref{aff115},\ref{aff6}}
\and J.~Bel\orcid{0009-0006-7837-1866}\inst{\ref{aff137}}
\and J.~R.~Bermejo-Climent\inst{\ref{aff138},\ref{aff139}}
\and D.~Bertacca\orcid{0000-0002-2490-7139}\inst{\ref{aff104},\ref{aff14},\ref{aff60}}
\and M.~Bethermin\orcid{0000-0002-3915-2015}\inst{\ref{aff140}}
\and F.~Beutler\orcid{0000-0003-0467-5438}\inst{\ref{aff47}}
\and M.~Bonici\orcid{0000-0002-8430-126X}\inst{\ref{aff98},\ref{aff37}}
\and F.~Bouch\`e\orcid{0000-0002-4663-1786}\inst{\ref{aff141},\ref{aff43}}
\and M.~L.~Brown\orcid{0000-0002-0370-8077}\inst{\ref{aff48}}
\and S.~Bruton\orcid{0000-0002-6503-5218}\inst{\ref{aff13}}
\and A.~Calabro\orcid{0000-0003-2536-1614}\inst{\ref{aff9}}
\and B.~Camacho~Quevedo\orcid{0000-0002-8789-4232}\inst{\ref{aff19},\ref{aff21},\ref{aff2}}
\and C.~S.~Carvalho\inst{\ref{aff142}}
\and T.~E.~Collett\orcid{0000-0001-5564-3140}\inst{\ref{aff128}}
\and A.~R.~Cooray\orcid{0000-0002-3892-0190}\inst{\ref{aff143}}
\and P.~Corcho-Caballero\orcid{0000-0001-6327-7080}\inst{\ref{aff10}}
\and B.~Csizi\orcid{0000-0003-3227-6581}\inst{\ref{aff110}}
\and O.~Cucciati\orcid{0000-0002-9336-7551}\inst{\ref{aff6}}
\and T.~de~Boer\orcid{0000-0001-5486-2747}\inst{\ref{aff44}}
\and G.~Desprez\orcid{0000-0001-8325-1742}\inst{\ref{aff10}}
\and A.~D\'iaz-S\'anchez\orcid{0000-0003-0748-4768}\inst{\ref{aff144}}
\and S.~Di~Domizio\orcid{0000-0003-2863-5895}\inst{\ref{aff27},\ref{aff28}}
\and J.~M.~Diego\orcid{0000-0001-9065-3926}\inst{\ref{aff145}}
\and P.-A.~Duc\orcid{0000-0003-3343-6284}\inst{\ref{aff140}}
\and V.~Duret\orcid{0009-0009-0383-4960}\inst{\ref{aff146}}
\and A.~Enia\orcid{0000-0002-0200-2857}\inst{\ref{aff121},\ref{aff6}}
\and Y.~Fang\orcid{0000-0002-0334-6950}\inst{\ref{aff26}}
\and A.~Farina\orcid{0009-0000-3420-929X}\inst{\ref{aff17},\ref{aff28}}
\and A.~M.~N.~Ferguson\orcid{0000-0001-7934-1278}\inst{\ref{aff47}}
\and A.~Finoguenov\orcid{0000-0002-4606-5403}\inst{\ref{aff78}}
\and A.~Fontana\orcid{0000-0003-3820-2823}\inst{\ref{aff9}}
\and A.~Franco\orcid{0000-0002-4761-366X}\inst{\ref{aff147},\ref{aff148},\ref{aff149}}
\and N.~Frusciante\orcid{0000-0002-7375-1230}\inst{\ref{aff150},\ref{aff151}}
\and A.~Fumagalli\orcid{0009-0004-0300-2535}\inst{\ref{aff2}}
\and K.~Ganga\orcid{0000-0001-8159-8208}\inst{\ref{aff90}}
\and E.~Gaztanaga\orcid{0000-0001-9632-0815}\inst{\ref{aff41},\ref{aff42},\ref{aff128}}
\and Z.~Ghaffari\orcid{0000-0002-6467-8078}\inst{\ref{aff2},\ref{aff19}}
\and F.~Giacomini\orcid{0000-0002-3129-2814}\inst{\ref{aff23}}
\and F.~Gianotti\orcid{0000-0003-4666-119X}\inst{\ref{aff6}}
\and G.~Gozaliasl\orcid{0000-0002-0236-919X}\inst{\ref{aff152},\ref{aff78}}
\and A.~Gruppuso\orcid{0000-0001-9272-5292}\inst{\ref{aff6},\ref{aff23}}
\and M.~Guidi\orcid{0000-0001-9408-1101}\inst{\ref{aff22},\ref{aff6}}
\and N.~A.~Hatch\orcid{0000-0001-5600-0534}\inst{\ref{aff153}}
\and C.~Hern\'andez-Monteagudo\orcid{0000-0001-5471-9166}\inst{\ref{aff122},\ref{aff46}}
\and T.~Heuberger\orcid{0009-0009-3474-8467}\inst{\ref{aff84}}
\and J.~Hjorth\orcid{0000-0002-4571-2306}\inst{\ref{aff154}}
\and M.~Jauzac\orcid{0000-0003-1974-8732}\inst{\ref{aff123},\ref{aff87},\ref{aff155},\ref{aff156},\ref{aff106}}
\and S.~Joudaki\orcid{0000-0001-8820-673X}\inst{\ref{aff39}}
\and J.~J.~E.~Kajava\orcid{0000-0002-3010-8333}\inst{\ref{aff157},\ref{aff158},\ref{aff159}}
\and D.~Karagiannis\orcid{0000-0002-4927-0816}\inst{\ref{aff160},\ref{aff115},\ref{aff161}}
\and J.~Kim\orcid{0000-0003-2776-2761}\inst{\ref{aff64}}
\and C.~C.~Kirkpatrick\inst{\ref{aff162}}
\and A.~Kov\'acs\orcid{0000-0002-5825-579X}\inst{\ref{aff139},\ref{aff138}}
\and I.~Kova{\v{c}}i{\'{c}}\orcid{0000-0001-6751-3263}\inst{\ref{aff136}}
\and K.~Koyama\orcid{0000-0001-6727-6915}\inst{\ref{aff128}}
\and S.~Kruk\orcid{0000-0001-8010-8879}\inst{\ref{aff16}}
\and M.~C.~Lam\orcid{0000-0002-9347-2298}\inst{\ref{aff47}}
\and M.~Lattanzi\orcid{0000-0003-1059-2532}\inst{\ref{aff116}}
\and S.~Lee\orcid{0000-0002-8289-740X}\inst{\ref{aff163},\ref{aff69}}
\and J.~Le~Graet\orcid{0000-0001-6523-7971}\inst{\ref{aff11}}
\and M.~Lembo\orcid{0000-0002-5271-5070}\inst{\ref{aff88}}
\and M.~D.~Lepinzan\orcid{0000-0003-1287-9801}\inst{\ref{aff2}}
\and J.~Lesgourgues\orcid{0000-0001-7627-353X}\inst{\ref{aff164}}
\and T.~I.~Liaudat\orcid{0000-0002-9104-314X}\inst{\ref{aff165}}
\and L.~Linke\orcid{0000-0002-2622-8113}\inst{\ref{aff110}}
\and S.~J.~Liu\orcid{0000-0001-7680-2139}\inst{\ref{aff61}}
\and F.~Mannucci\orcid{0000-0002-4803-2381}\inst{\ref{aff166}}
\and F.~R.~Marleau\orcid{0000-0002-1442-2947}\inst{\ref{aff110}}
\and C.~J.~A.~P.~Martins\orcid{0000-0002-4886-9261}\inst{\ref{aff167},\ref{aff32}}
\and M.~Migliaccio\inst{\ref{aff168},\ref{aff169}}
\and M.~Miluzio\inst{\ref{aff16},\ref{aff170}}
\and G.~Morgante\inst{\ref{aff6}}
\and K.~Naidoo\orcid{0000-0002-9182-1802}\inst{\ref{aff128},\ref{aff12}}
\and P.~Natoli\orcid{0000-0003-0126-9100}\inst{\ref{aff115},\ref{aff116}}
\and A.~Navarro-Alsina\orcid{0000-0002-3173-2592}\inst{\ref{aff85}}
\and A.~Nersesian\orcid{0000-0001-6843-409X}\inst{\ref{aff171},\ref{aff136}}
\and S.~Nesseris\orcid{0000-0002-0567-0324}\inst{\ref{aff105}}
\and L.~Nicastro\orcid{0000-0001-8534-6788}\inst{\ref{aff6}}
\and F.~Oppizzi\orcid{0000-0003-3904-8370}\inst{\ref{aff28},\ref{aff60},\ref{aff104}}
\and F.~Pace\orcid{0000-0001-8039-0480}\inst{\ref{aff34},\ref{aff35},\ref{aff36},\ref{aff56}}
\and M.~J.~Page\orcid{0000-0002-6689-6271}\inst{\ref{aff54}}
\and D.~Paoletti\orcid{0000-0003-4761-6147}\inst{\ref{aff6},\ref{aff62}}
\and F.~Passalacqua\orcid{0000-0002-8606-4093}\inst{\ref{aff60}}
\and C.~Pattison\orcid{0000-0003-3272-2617}\inst{\ref{aff128}}
\and A.~Pisani\orcid{0000-0002-6146-4437}\inst{\ref{aff11}}
\and L.~Pizzuti\orcid{0000-0001-5654-7580}\inst{\ref{aff172}}
\and J.~J.~Popp\orcid{0000-0002-3724-1727}\inst{\ref{aff173}}
\and D.~Potter\orcid{0000-0002-0757-5195}\inst{\ref{aff174}}
\and M.~Poulain\orcid{0000-0002-7664-4510}\inst{\ref{aff175}}
\and G.~W.~Pratt\inst{\ref{aff18}}
\and S.~Quai\orcid{0000-0002-0449-8163}\inst{\ref{aff86},\ref{aff6}}
\and F.~Ricci\orcid{0000-0001-5742-5980}\inst{\ref{aff176},\ref{aff9}}
\and W.~Roster\orcid{0000-0002-9149-6528}\inst{\ref{aff25}}
\and M.~Sahl\'en\orcid{0000-0003-0973-4804}\inst{\ref{aff177}}
\and D.~B.~Sanders\orcid{0000-0002-1233-9998}\inst{\ref{aff44}}
\and A.~Schneider\orcid{0000-0001-7055-8104}\inst{\ref{aff174}}
\and D.~Sciotti\orcid{0009-0008-4519-2620}\inst{\ref{aff9},\ref{aff38}}
\and E.~Sellentin\orcid{0009-0002-2655-3458}\inst{\ref{aff178},\ref{aff4}}
\and S.~Serjeant\orcid{0000-0002-0517-7943}\inst{\ref{aff173}}
\and L.~C.~Smith\orcid{0000-0002-3259-2771}\inst{\ref{aff179}}
\and I.~Szapudi\orcid{0000-0003-2274-0301}\inst{\ref{aff44}}
\and M.~Talia\orcid{0000-0003-4352-2063}\inst{\ref{aff86},\ref{aff6}}
\and G.~Testera\orcid{0000-0003-2970-766X}\inst{\ref{aff28}}
\and R.~Teyssier\orcid{0000-0001-7689-0933}\inst{\ref{aff180}}
\and S.~Tosi\orcid{0000-0002-7275-9193}\inst{\ref{aff27},\ref{aff17},\ref{aff28}}
\and C.~Uhlemann\orcid{0000-0001-7831-1579}\inst{\ref{aff181}}
\and M.~Urbano\orcid{0000-0001-5640-0650}\inst{\ref{aff153},\ref{aff140}}
\and C.~Valieri\inst{\ref{aff23}}
\and A.~Venhola\orcid{0000-0001-6071-4564}\inst{\ref{aff175}}
\and D.~Vergani\orcid{0000-0003-0898-2216}\inst{\ref{aff6}}
\and A.~Viitanen\orcid{0000-0001-9383-786X}\inst{\ref{aff57},\ref{aff162},\ref{aff9}}
\and E.~Vilenius\orcid{0000-0002-6184-7681}\inst{\ref{aff54}}
\and S.~Vinciguerra\orcid{0009-0005-4018-3184}\inst{\ref{aff81}}
\and M.~von~Wietersheim-Kramsta\orcid{0000-0003-4986-5091}\inst{\ref{aff87},\ref{aff123}}
\and N.~A.~Walton\orcid{0000-0003-3983-8778}\inst{\ref{aff179}}
\and L.~Wang\orcid{0000-0002-6736-9158}\inst{\ref{aff182},\ref{aff10}}
\and A.~H.~Wright\orcid{0000-0001-7363-7932}\inst{\ref{aff183}}
\and D.~Scott\orcid{0000-0002-6878-9840}\inst{\ref{aff184}}}
										   
\institute{Space Science Data Center, Italian Space Agency, via del Politecnico snc, 00133 Roma, Italy\label{aff1}
\and
INAF-Osservatorio Astronomico di Trieste, Via G. B. Tiepolo 11, 34143 Trieste, Italy\label{aff2}
\and
Caltech/IPAC, 1200 E. California Blvd., Pasadena, CA 91125, USA\label{aff3}
\and
Leiden Observatory, Leiden University, Einsteinweg 55, 2333 CC Leiden, The Netherlands\label{aff4}
\and
Aerospace Logistics Technology Engineering Company, Corso Marche 79, 10146 Torino, Italy\label{aff5}
\and
INAF-Osservatorio di Astrofisica e Scienza dello Spazio di Bologna, Via Piero Gobetti 93/3, 40129 Bologna, Italy\label{aff6}
\and
Centre de Recherche Astrophysique de Lyon, UMR5574, CNRS, Universit\'e Claude Bernard Lyon 1, ENS de Lyon, 69230, Saint-Genis-Laval, France\label{aff7}
\and
Universit\'e Lyon 1, CNRS/IN2P3, IP2I Lyon, Villeurbanne, France\label{aff8}
\and
INAF-Osservatorio Astronomico di Roma, Via Frascati 33, 00078 Monteporzio Catone, Italy\label{aff9}
\and
Kapteyn Astronomical Institute, University of Groningen, PO Box 800, 9700 AV Groningen, The Netherlands\label{aff10}
\and
Aix-Marseille Universit\'e, CNRS/IN2P3, CPPM, Marseille, France\label{aff11}
\and
Max-Planck-Institut f\"ur Astronomie, K\"onigstuhl 17, 69117 Heidelberg, Germany\label{aff12}
\and
California Institute of Technology, 1200 E California Blvd, Pasadena, CA 91125, USA\label{aff13}
\and
INAF-Osservatorio Astronomico di Padova, Via dell'Osservatorio 5, 35122 Padova, Italy\label{aff14}
\and
Infrared Processing and Analysis Center, California Institute of Technology, Pasadena, CA 91125, USA\label{aff15}
\and
ESAC/ESA, Camino Bajo del Castillo, s/n., Urb. Villafranca del Castillo, 28692 Villanueva de la Ca\~nada, Madrid, Spain\label{aff16}
\and
INAF-Osservatorio Astronomico di Brera, Via Brera 28, 20122 Milano, Italy\label{aff17}
\and
Universit\'e Paris-Saclay, Universit\'e Paris Cit\'e, CEA, CNRS, AIM, 91191, Gif-sur-Yvette, France\label{aff18}
\and
IFPU, Institute for Fundamental Physics of the Universe, via Beirut 2, 34151 Trieste, Italy\label{aff19}
\and
INFN, Sezione di Trieste, Via Valerio 2, 34127 Trieste TS, Italy\label{aff20}
\and
SISSA, International School for Advanced Studies, Via Bonomea 265, 34136 Trieste TS, Italy\label{aff21}
\and
Dipartimento di Fisica e Astronomia, Universit\`a di Bologna, Via Gobetti 93/2, 40129 Bologna, Italy\label{aff22}
\and
INFN-Sezione di Bologna, Viale Berti Pichat 6/2, 40127 Bologna, Italy\label{aff23}
\and
ATG Europe BV, Huygensstraat 34, 2201 DK Noordwijk, The Netherlands\label{aff24}
\and
Max Planck Institute for Extraterrestrial Physics, Giessenbachstr. 1, 85748 Garching, Germany\label{aff25}
\and
Universit\"ats-Sternwarte M\"unchen, Fakult\"at f\"ur Physik, Ludwig-Maximilians-Universit\"at M\"unchen, Scheinerstr.~1, 81679 M\"unchen, Germany\label{aff26}
\and
Dipartimento di Fisica, Universit\`a di Genova, Via Dodecaneso 33, 16146, Genova, Italy\label{aff27}
\and
INFN-Sezione di Genova, Via Dodecaneso 33, 16146, Genova, Italy\label{aff28}
\and
Department of Physics "E. Pancini", University Federico II, Via Cinthia 6, 80126, Napoli, Italy\label{aff29}
\and
INAF-Osservatorio Astronomico di Capodimonte, Via Moiariello 16, 80131 Napoli, Italy\label{aff30}
\and
European Southern Observatory, Karl-Schwarzschild-Str.~2, 85748 Garching, Germany\label{aff31}
\and
Instituto de Astrof\'isica e Ci\^encias do Espa\c{c}o, Universidade do Porto, CAUP, Rua das Estrelas, PT4150-762 Porto, Portugal\label{aff32}
\and
Faculdade de Ci\^encias da Universidade do Porto, Rua do Campo de Alegre, 4150-007 Porto, Portugal\label{aff33}
\and
Dipartimento di Fisica, Universit\`a degli Studi di Torino, Via P. Giuria 1, 10125 Torino, Italy\label{aff34}
\and
INFN-Sezione di Torino, Via P. Giuria 1, 10125 Torino, Italy\label{aff35}
\and
INAF-Osservatorio Astrofisico di Torino, Via Osservatorio 20, 10025 Pino Torinese (TO), Italy\label{aff36}
\and
INAF-IASF Milano, Via Alfonso Corti 12, 20133 Milano, Italy\label{aff37}
\and
INFN-Sezione di Roma, Piazzale Aldo Moro, 2 - c/o Dipartimento di Fisica, Edificio G. Marconi, 00185 Roma, Italy\label{aff38}
\and
Centro de Investigaciones Energ\'eticas, Medioambientales y Tecnol\'ogicas (CIEMAT), Avenida Complutense 40, 28040 Madrid, Spain\label{aff39}
\and
Port d'Informaci\'{o} Cient\'{i}fica, Campus UAB, C. Albareda s/n, 08193 Bellaterra (Barcelona), Spain\label{aff40}
\and
Institute of Space Sciences (ICE, CSIC), Campus UAB, Carrer de Can Magrans, s/n, 08193 Barcelona, Spain\label{aff41}
\and
Institut d'Estudis Espacials de Catalunya (IEEC),  Edifici RDIT, Campus UPC, 08860 Castelldefels, Barcelona, Spain\label{aff42}
\and
INFN -- Sezione di Napoli, Via Cinthia 6, 80126, Napoli, Italy\label{aff43}
\and
Institute for Astronomy, University of Hawaii, 2680 Woodlawn Drive, Honolulu, HI 96822, USA\label{aff44}
\and
Dipartimento di Fisica e Astronomia "Augusto Righi" - Alma Mater Studiorum Universit\`a di Bologna, Viale Berti Pichat 6/2, 40127 Bologna, Italy\label{aff45}
\and
Instituto de Astrof\'{\i}sica de Canarias, E-38205 La Laguna, Tenerife, Spain\label{aff46}
\and
Institute for Astronomy, University of Edinburgh, Royal Observatory, Blackford Hill, Edinburgh EH9 3HJ, UK\label{aff47}
\and
Jodrell Bank Centre for Astrophysics, Department of Physics and Astronomy, University of Manchester, Oxford Road, Manchester M13 9PL, UK\label{aff48}
\and
European Space Agency/ESRIN, Largo Galileo Galilei 1, 00044 Frascati, Roma, Italy\label{aff49}
\and
Institut de Ci\`{e}ncies del Cosmos (ICCUB), Universitat de Barcelona (IEEC-UB), Mart\'{i} i Franqu\`{e}s 1, 08028 Barcelona, Spain\label{aff50}
\and
Instituci\'o Catalana de Recerca i Estudis Avan\c{c}ats (ICREA), Passeig de Llu\'{\i}s Companys 23, 08010 Barcelona, Spain\label{aff51}
\and
Institut de Ciencies de l'Espai (IEEC-CSIC), Campus UAB, Carrer de Can Magrans, s/n Cerdanyola del Vall\'es, 08193 Barcelona, Spain\label{aff52}
\and
Universit\'e Lyon 1, CNRS/IN2P3, IUF, IP2I Lyon, Villeurbanne, France \label{aff53}
\and
Mullard Space Science Laboratory, University College London, Holmbury St Mary, Dorking, Surrey RH5 6NT, UK\label{aff54}
\and
Departamento de F\'isica, Faculdade de Ci\^encias, Universidade de Lisboa, Edif\'icio C8, Campo Grande, PT1749-016 Lisboa, Portugal\label{aff55}
\and
Instituto de Astrof\'isica e Ci\^encias do Espa\c{c}o, Faculdade de Ci\^encias, Universidade de Lisboa, Campo Grande, 1749-016 Lisboa, Portugal\label{aff56}
\and
Department of Astronomy, University of Geneva, ch. d'Ecogia 16, 1290 Versoix, Switzerland\label{aff57}
\and
Universit\'e Paris-Saclay, CNRS, Institut d'astrophysique spatiale, 91405, Orsay, France\label{aff58}
\and
Institut universitaire de France (IUF), 1 rue Descartes, 75231 PARIS CEDEX 05, France\label{aff59}
\and
INFN-Padova, Via Marzolo 8, 35131 Padova, Italy\label{aff60}
\and
INAF-Istituto di Astrofisica e Planetologia Spaziali, via del Fosso del Cavaliere, 100, 00100 Roma, Italy\label{aff61}
\and
INFN-Bologna, Via Irnerio 46, 40126 Bologna, Italy\label{aff62}
\and
School of Physics, HH Wills Physics Laboratory, University of Bristol, Tyndall Avenue, Bristol, BS8 1TL, UK\label{aff63}
\and
Department of Physics, University of Oxford, Keble Road, Oxford OX1 3RH, UK\label{aff64}
\and
University Observatory, LMU Faculty of Physics, Scheinerstr.~1, 81679 Munich, Germany\label{aff65}
\and
Dipartimento di Fisica "Aldo Pontremoli", Universit\`a degli Studi di Milano, Via Celoria 16, 20133 Milano, Italy\label{aff66}
\and
INFN-Sezione di Milano, Via Celoria 16, 20133 Milano, Italy\label{aff67}
\and
Institute of Theoretical Astrophysics, University of Oslo, P.O. Box 1029 Blindern, 0315 Oslo, Norway\label{aff68}
\and
Jet Propulsion Laboratory, California Institute of Technology, 4800 Oak Grove Drive, Pasadena, CA, 91109, USA\label{aff69}
\and
School of Physics and Astronomy, Lancaster University, Lancaster, LA1 4YB, UK\label{aff70}
\and
Felix Hormuth Engineering, Goethestr. 17, 69181 Leimen, Germany\label{aff71}
\and
Technical University of Denmark, Elektrovej 327, 2800 Kgs. Lyngby, Denmark\label{aff72}
\and
Cosmic Dawn Center (DAWN), Denmark\label{aff73}
\and
Universit\'e PSL, Observatoire de Paris, Sorbonne Universit\'e, CNRS, LERMA, 75014, Paris, France\label{aff74}
\and
Universit\'e Paris-Cit\'e, 5 Rue Thomas Mann, 75013, Paris, France\label{aff75}
\and
NASA Goddard Space Flight Center, Greenbelt, MD 20771, USA\label{aff76}
\and
Universit\'e de Gen\`eve, D\'epartement de Physique Th\'eorique and Centre for Astroparticle Physics, 24 quai Ernest-Ansermet, CH-1211 Gen\`eve 4, Switzerland\label{aff77}
\and
Department of Physics, P.O. Box 64, University of Helsinki, 00014 Helsinki, Finland\label{aff78}
\and
Helsinki Institute of Physics, Gustaf H{\"a}llstr{\"o}min katu 2, University of Helsinki, 00014 Helsinki, Finland\label{aff79}
\and
Laboratoire d'etude de l'Univers et des phenomenes eXtremes, Observatoire de Paris, Universit\'e PSL, Sorbonne Universit\'e, CNRS, 92190 Meudon, France\label{aff80}
\and
Aix-Marseille Universit\'e, CNRS, CNES, LAM, Marseille, France\label{aff81}
\and
SKAO, Jodrell Bank, Lower Withington, Macclesfield SK11 9FT, UK\label{aff82}
\and
Centre de Calcul de l'IN2P3/CNRS, 21 avenue Pierre de Coubertin 69627 Villeurbanne Cedex, France\label{aff83}
\and
University of Applied Sciences and Arts of Northwestern Switzerland, School of Computer Science, 5210 Windisch, Switzerland\label{aff84}
\and
Universit\"at Bonn, Argelander-Institut f\"ur Astronomie, Auf dem H\"ugel 71, 53121 Bonn, Germany\label{aff85}
\and
Dipartimento di Fisica e Astronomia "Augusto Righi" - Alma Mater Studiorum Universit\`a di Bologna, via Piero Gobetti 93/2, 40129 Bologna, Italy\label{aff86}
\and
Department of Physics, Institute for Computational Cosmology, Durham University, South Road, Durham, DH1 3LE, UK\label{aff87}
\and
Institut d'Astrophysique de Paris, UMR 7095, CNRS, and Sorbonne Universit\'e, 98 bis boulevard Arago, 75014 Paris, France\label{aff88}
\and
Niels Bohr Institute, University of Copenhagen, Jagtvej 128, 2200 Copenhagen, Denmark\label{aff89}
\and
Universit\'e Paris Cit\'e, CNRS, Astroparticule et Cosmologie, 75013 Paris, France\label{aff90}
\and
CNRS-UCB International Research Laboratory, Centre Pierre Bin\'etruy, IRL2007, CPB-IN2P3, Berkeley, USA\label{aff91}
\and
Institute of Physics, Laboratory of Astrophysics, Ecole Polytechnique F\'ed\'erale de Lausanne (EPFL), Observatoire de Sauverny, 1290 Versoix, Switzerland\label{aff92}
\and
Dipartimento di Fisica - Sezione di Astronomia, Universit\`a di Trieste, Via Tiepolo 11, 34131 Trieste, Italy\label{aff93}
\and
Telespazio UK S.L. for European Space Agency (ESA), Camino bajo del Castillo, s/n, Urbanizacion Villafranca del Castillo, Villanueva de la Ca\~nada, 28692 Madrid, Spain\label{aff94}
\and
Institut de F\'{i}sica d'Altes Energies (IFAE), The Barcelona Institute of Science and Technology, Campus UAB, 08193 Bellaterra (Barcelona), Spain\label{aff95}
\and
School of Mathematics and Physics, University of Surrey, Guildford, Surrey, GU2 7XH, UK\label{aff96}
\and
European Space Agency/ESTEC, Keplerlaan 1, 2201 AZ Noordwijk, The Netherlands\label{aff97}
\and
Waterloo Centre for Astrophysics, University of Waterloo, Waterloo, Ontario N2L 3G1, Canada\label{aff98}
\and
Department of Physics and Astronomy, University of Waterloo, Waterloo, Ontario N2L 3G1, Canada\label{aff99}
\and
Perimeter Institute for Theoretical Physics, Waterloo, Ontario N2L 2Y5, Canada\label{aff100}
\and
Centre National d'Etudes Spatiales -- Centre spatial de Toulouse, 18 avenue Edouard Belin, 31401 Toulouse Cedex 9, France\label{aff101}
\and
Institute of Space Science, Str. Atomistilor, nr. 409 M\u{a}gurele, Ilfov, 077125, Romania\label{aff102}
\and
Institute for Particle Physics and Astrophysics, Dept. of Physics, ETH Zurich, Wolfgang-Pauli-Strasse 27, 8093 Zurich, Switzerland\label{aff103}
\and
Dipartimento di Fisica e Astronomia "G. Galilei", Universit\`a di Padova, Via Marzolo 8, 35131 Padova, Italy\label{aff104}
\and
Instituto de F\'isica Te\'orica UAM-CSIC, Campus de Cantoblanco, 28049 Madrid, Spain\label{aff105}
\and
Institut de Recherche en Astrophysique et Plan\'etologie (IRAP), Universit\'e de Toulouse, CNRS, UPS, CNES, 14 Av. Edouard Belin, 31400 Toulouse, France\label{aff106}
\and
Universit\'e St Joseph; Faculty of Sciences, Beirut, Lebanon\label{aff107}
\and
Departamento de F\'isica, FCFM, Universidad de Chile, Blanco Encalada 2008, Santiago, Chile\label{aff108}
\and
Universit\'e C\^{o}te d'Azur, Observatoire de la C\^{o}te d'Azur, CNRS, Laboratoire Lagrange, Bd de l'Observatoire, CS 34229, 06304 Nice cedex 4, France\label{aff109}
\and
Universit\"at Innsbruck, Institut f\"ur Astro- und Teilchenphysik, Technikerstr. 25/8, 6020 Innsbruck, Austria\label{aff110}
\and
Cosmic Dawn Center (DAWN)\label{aff111}
\and
Universidad Polit\'ecnica de Cartagena, Departamento de Electr\'onica y Tecnolog\'ia de Computadoras,  Plaza del Hospital 1, 30202 Cartagena, Spain\label{aff112}
\and
European University of Technology EUt+, European Union\label{aff113}
\and
Centre for Information Technology, University of Groningen, P.O. Box 11044, 9700 CA Groningen, The Netherlands\label{aff114}
\and
Dipartimento di Fisica e Scienze della Terra, Universit\`a degli Studi di Ferrara, Via Giuseppe Saragat 1, 44122 Ferrara, Italy\label{aff115}
\and
Istituto Nazionale di Fisica Nucleare, Sezione di Ferrara, Via Giuseppe Saragat 1, 44122 Ferrara, Italy\label{aff116}
\and
ICSC - Centro Nazionale di Ricerca in High Performance Computing, Big Data e Quantum Computing, Via Magnanelli 2, Bologna, Italy\label{aff117}
\and
INAF, Istituto di Radioastronomia, Via Piero Gobetti 101, 40129 Bologna, Italy\label{aff118}
\and
Astronomical Observatory of the Autonomous Region of the Aosta Valley (OAVdA), Loc. Lignan 39, I-11020, Nus (Aosta Valley), Italy\label{aff119}
\and
Department of Mathematical Physics, Institute of Physics, University of S\~ao Paulo, R. do Mat\~ao 1371, 05508-090, S\~ao Paulo, SP, Brazil\label{aff120}
\and
CEA Saclay, DFR/IRFU, Service d'Astrophysique, Bat. 709, 91191 Gif-sur-Yvette, France\label{aff121}
\and
Universidad de La Laguna, Dpto. Astrof\'\i sica, E-38206 La Laguna, Tenerife, Spain\label{aff122}
\and
Department of Physics, Centre for Extragalactic Astronomy, Durham University, South Road, Durham, DH1 3LE, UK\label{aff123}
\and
Oskar Klein Centre for Cosmoparticle Physics, Department of Physics, Stockholm University, Stockholm, SE-106 91, Sweden\label{aff124}
\and
Astrophysics Group, Blackett Laboratory, Imperial College London, London SW7 2AZ, UK\label{aff125}
\and
Univ. Grenoble Alpes, CNRS, Grenoble INP, LPSC-IN2P3, 53, Avenue des Martyrs, 38000 Grenoble, France\label{aff126}
\and
Dipartimento di Fisica, Sapienza Universit\`a di Roma, Piazzale Aldo Moro 2, 00185 Roma, Italy\label{aff127}
\and
Institute of Cosmology and Gravitation, University of Portsmouth, Portsmouth PO1 3FX, UK\label{aff128}
\and
Institut d'Astrophysique de Paris, 98bis Boulevard Arago, 75014, Paris, France\label{aff129}
\and
ICL, Junia, Universit\'e Catholique de Lille, LITL, 59000 Lille, France\label{aff130}
\and
Center for Astrophysics and Cosmology, University of Nova Gorica, Nova Gorica, Slovenia\label{aff131}
\and
Department of Astronomy, Indiana University, 727 East Third Street, Bloomington, IN 47405, USA\label{aff132}
\and
CERCA/ISO, Department of Physics, Case Western Reserve University, 10900 Euclid Avenue, Cleveland, OH 44106, USA\label{aff133}
\and
Laboratoire Univers et Th\'eorie, Observatoire de Paris, Universit\'e PSL, Universit\'e Paris Cit\'e, CNRS, 92190 Meudon, France\label{aff134}
\and
Departamento de F{\'\i}sica Fundamental. Universidad de Salamanca. Plaza de la Merced s/n. 37008 Salamanca, Spain\label{aff135}
\and
Universiteit Gent, Department of Physics and Astronomy, Proeftuinstraat 86 N3, 9000 Ghent, Belgium
\label{aff136}
\and
Aix-Marseille Universit\'e, Universit\'e de Toulon, CNRS, CPT, Marseille, France\label{aff137}
\and
Konkoly Observatory, HUN-REN CSFK, MTA Centre of Excellence, Budapest, Konkoly Thege Mikl\'os {\'u}t 15-17. H-1121, Hungary\label{aff138}
\and
MTA-CSFK Lend\"ulet Large-Scale Structure Research Group, Konkoly-Thege Mikl\'os \'ut 15-17, H-1121 Budapest, Hungary\label{aff139}
\and
Universit\'e de Strasbourg, CNRS, Observatoire astronomique de Strasbourg, UMR 7550, 67000 Strasbourg, France\label{aff140}
\and
Scuola Superiore Meridionale, Via Mezzocannone 4, 80138, Napoli, Italy\label{aff141}
\and
Instituto de Astrof\'isica e Ci\^encias do Espa\c{c}o, Faculdade de Ci\^encias, Universidade de Lisboa, Tapada da Ajuda, 1349-018 Lisboa, Portugal\label{aff142}
\and
Department of Physics \& Astronomy, University of California Irvine, Irvine CA 92697, USA\label{aff143}
\and
Departamento F\'isica Aplicada, Universidad Polit\'ecnica de Cartagena, Campus Muralla del Mar, 30202 Cartagena, Murcia, Spain\label{aff144}
\and
Instituto de F\'isica de Cantabria, Edificio Juan Jord\'a, Avenida de los Castros, 39005 Santander, Spain\label{aff145}
\and
Univ. Grenoble Alpes, CNRS, IPAG, 38000 Grenoble, France\label{aff146}
\and
Department of Mathematics and Physics E. De Giorgi, University of Salento, Via per Arnesano, CP-I93, 73100, Lecce, Italy\label{aff147}
\and
INFN, Sezione di Lecce, Via per Arnesano, CP-193, 73100, Lecce, Italy\label{aff148}
\and
INAF-Sezione di Lecce, c/o Dipartimento Matematica e Fisica, Via per Arnesano, 73100, Lecce, Italy\label{aff149}
\and
Dipartimento di Fisica "E. Pancini", Universita degli Studi di Napoli Federico II, Via Cinthia 6, 80126, Napoli, Italy\label{aff150}
\and
INFN Sezione di Napoli, Universita degli Studi di Napoli "Federico II", Compl. Univ. di Monte S. Angelo, Edificio G, Via Cinthia, I-80126, Napoli, Italy\label{aff151}
\and
Department of Computer Science, Aalto University, PO Box 15400, Espoo, FI-00 076, Finland\label{aff152}
\and
School of Physics and Astronomy, University of Nottingham, University Park, Nottingham NG7 2RD, UK\label{aff153}
\and
DARK, Niels Bohr Institute, University of Copenhagen, Jagtvej 155, 2200 Copenhagen, Denmark\label{aff154}
\and
Astrophysics Research Centre, University of KwaZulu-Natal, Westville Campus, Durban 4041, South Africa\label{aff155}
\and
School of Mathematics, Statistics \& Computer Science, University of KwaZulu-Natal, Westville Campus, Durban 4041, South Africa\label{aff156}
\and
Department of Physics and Astronomy, Vesilinnantie 5, University of Turku, 20014 Turku, Finland\label{aff157}
\and
Finnish Centre for Astronomy with ESO (FINCA), Quantum, Vesilinnantie 5, University of Turku, 20014 Turku, Finland\label{aff158}
\and
Serco for European Space Agency (ESA), Camino bajo del Castillo, s/n, Urbanizacion Villafranca del Castillo, Villanueva de la Ca\~nada, 28692 Madrid, Spain\label{aff159}
\and
Van Swinderen Institute, University of Groningen, Nijenborgh 3, 9747 AG Groningen, The Netherlands\label{aff160}
\and
Department of Physics and Astronomy, University of the Western Cape, Bellville, Cape Town, 7535, South Africa\label{aff161}
\and
Department of Physics and Helsinki Institute of Physics, Gustaf H\"allstr\"omin katu 2, University of Helsinki, 00014 Helsinki, Finland\label{aff162}
\and
Ohio University, Physics \& Astronomy Department,1 Ohio University, Athens, OH 45701, USA\label{aff163}
\and
Institute for Theoretical Particle Physics and Cosmology (TTK), RWTH Aachen University, 52056 Aachen, Germany\label{aff164}
\and
IRFU, CEA, Universit\'e Paris-Saclay 91191 Gif-sur-Yvette Cedex, France\label{aff165}
\and
INAF-Osservatorio Astrofisico di Arcetri, Largo E. Fermi 5, 50125, Firenze, Italy\label{aff166}
\and
Centro de Astrof\'{\i}sica da Universidade do Porto, Rua das Estrelas, 4150-762 Porto, Portugal\label{aff167}
\and
Dipartimento di Fisica, Universit\`a di Roma Tor Vergata, Via della Ricerca Scientifica 1, Roma, Italy\label{aff168}
\and
INFN, Sezione di Roma 2, Via della Ricerca Scientifica 1, Roma, Italy\label{aff169}
\and
HE Space for European Space Agency (ESA), Camino bajo del Castillo, s/n, Urbanizacion Villafranca del Castillo, Villanueva de la Ca\~nada, 28692 Madrid, Spain\label{aff170}
\and
STAR Institute, University of Li{\`e}ge, Quartier Agora, All\'ee du six Ao\^ut 19c, 4000 Li\`ege, Belgium\label{aff171}
\and
Dipartimento di Fisica ``G. Occhialini", Universit\`a degli Studi di Milano Bicocca, Piazza della Scienza 3, 20126 Milano, Italy\label{aff172}
\and
School of Physical Sciences, The Open University, Milton Keynes, MK7 6AA, UK\label{aff173}
\and
Department of Astrophysics, University of Zurich, Winterthurerstrasse 190, 8057 Zurich, Switzerland\label{aff174}
\and
Space physics and astronomy research unit, University of Oulu, Pentti Kaiteran katu 1, FI-90014 Oulu, Finland\label{aff175}
\and
Department of Mathematics and Physics, Roma Tre University, Via della Vasca Navale 84, 00146 Rome, Italy\label{aff176}
\and
Theoretical astrophysics, Department of Physics and Astronomy, Uppsala University, Box 524, 751 20 Uppsala, Sweden\label{aff177}
\and
Mathematical Institute, University of Leiden, Einsteinweg 55, 2333 CA Leiden, The Netherlands\label{aff178}
\and
Institute of Astronomy, University of Cambridge, Madingley Road, Cambridge CB3 0HA, UK\label{aff179}
\and
Department of Astrophysical Sciences, Peyton Hall, Princeton University, Princeton, NJ 08544, USA\label{aff180}
\and
Fakult\"at f\"ur Physik, Universit\"at Bielefeld, Postfach 100131, 33501 Bielefeld, Germany\label{aff181}
\and
SRON Netherlands Institute for Space Research, Landleven 12, 9747 AD, Groningen, The Netherlands\label{aff182}
\and
Ruhr University Bochum, Faculty of Physics and Astronomy, Astronomical Institute (AIRUB), German Centre for Cosmological Lensing (GCCL), 44780 Bochum, Germany\label{aff183}
\and
Department of Physics and Astronomy, University of British Columbia, Vancouver, BC V6T 1Z1, Canada\label{aff184}
}

\date{}

\abstract 
{
The \Euclid\ \ac{NIR PF} converts the raw observations acquired by the \ac{NISP} into calibrated images and source-level products. Following the first implementation delivered with the Euclid Quick Data Release 1 (Q1), the \ac{NIR PF} has been revised for the first major Euclid Data Release (DR1), with improvements to detector-level processing, calibration, image reconstruction, and quality control. The DR1 processing includes observations covering more than $1900\,\rm{deg}^2$ and has produced 71\,659 calibrated \ac{NIR} images. We assess the performance of the processing chain through extensive astrometric and photometric validation. The astrometric solution reaches an average root-mean-square (RMS) of 10\,mas, remaining below 20\,mas for 99\% of the dithers in all three NISP bands. Temporal variations in the instrumental throughput are typically corrected to a level of 2\,mmag RMS, while the relative photometric accuracy reaches 0.9\% for sources brighter than 19\,mag in \YE, \JE, and \HE, and the residual spatial variations of the photometric calibration across the focal plane are at the level of approximately 10\,mmag for the entire sample. Despite some remaining limitations, the DR1 NIR imaging products are consistent with the required astrometric and photometric performance over the large area covered by this release.
}


 \titlerunning{NIR processing and data products in DR1}
 \authorrunning{Euclid Collaboration: G. Polenta et al.}

\maketitle

\section{Introduction}
\label{sec:intro}

The ESA \Euclid\ space mission, launched on 1 July 2023, is conducting a wide and deep survey of the extragalactic sky to investigate the nature of dark matter and dark energy and the evolution of the large-scale structure of the Universe \citep{EuclidSkyOverview}. \Euclid\ is equipped with two complementary instruments: the visible imager \citep[VIS,][]{EuclidSkyVIS}; and the Near-Infrared Spectrometer and Photometer \cite[NISP,][]{EuclidSkyNISP}, which provides both near-infrared imaging and slitless spectroscopy. The NISP photometric channel covers the wavelength range from 950--2021\,nm through the three broad-band filters \YE, \JE, and \HE \citep{Schirmer-EP18}, using 16 HAWAII-2RG (H2RG) near-infrared detectors of 2048\,$\times$\,2048 pixels each, with a pixel scale of \ang{;;0.3} and a field of view of 0.57\,${\rm deg}^2$. 

The \ac{NIR PF} is the Euclid \ac{SGS} block \citep{DR1-TP033} responsible for transforming the raw NISP photometric observations into calibrated imaging products through a sequence of detector-level corrections, calibration, and image-processing steps, ultimately producing calibrated images and associated source-level products. The first implementation of the \ac{NIR PF} was described in the context of the Euclid Quick Data Release 1 \cite[Q1;][hereafter \citetalias{Q1-TP003}]{Q1-TP003}, which provided NISP imaging over 63.1\,${\rm deg}^2$ and served as the first large-scale validation of the processing chain. Building on this experience, the \ac{NIR PF} has been revised and extended for the first major Euclid Data Release \citep[DR1;][]{DR1cite,DR1-TP001}. The DR1 processing includes observations acquired between 1 October 2023 and 21 May 2025, covering a total area of more than 1900\,${\rm deg}^2$ across the \ac{EWS} and the \ac{EDS}, as well as the COSMOS field and dedicated calibration fields \citep{Scaramella-EP1, DR1-TP027}. In total, the DR1 processing produced $71\,659$ calibrated NIR images.

This paper presents the DR1 implementation of the \ac{NIR PF}, describing its updated processing algorithms, calibration strategy, data products, and validation on the first year of \Euclid\ survey observations. Specific aspects of the NIR data processing and characterisation are presented in dedicated companion papers, focusing on the astrometric calibration in \citet{DR1-TP004}, the \ac{PSF} in \citet{DR1-TP005}, and persistence effects in \citet{DR1-TP038}.

\section{In-flight stability of the NISP Instrument}
\label{sec:stability}

Overall, NISP has been operating very consistently since launch, with no major evolution of instrumental properties except the slow buildup of water ice on NISP's detector surfaces \citep{Schirmer-EP29}.
The physical number of unusable `bad' pixels has not changed significantly since launch, although their characterisation and flagging has been updated (see Sect.~\ref{sec:badpix}, and compare \cref{table:badpixdarks} for DR1 with table~1 in \citetalias{Q1-TP003} for Q1). The characterisation of charge persistence continues to be improved \cite[\cref{sec:persistence} here, and][]{DR1-TP038}, but the overall persistence properties of the detectors as a function of short- and long-term illumination do not seem to evolve. Similarly, focal plane geometry (relative positions and orientations of the detectors) derived from astrometric calibration is consistent with pre-launch metrology, thus indicating that the instrument itself is absolutely stable in this respect.

Remarkably, since launch and through Solar maximum, the flux of the \acp{LED} in NISP's calibration lamp \citep[NI-CU,][]{EuclidSkyNISPCU} have been stable to within 0.1\% at a given driver current. This order of stability was guaranteed for periods of 1200\,s only. Since the \acp{LED} illuminate the detectors directly without intervening optics, this stability allows accurate long-term monitoring of detector characteristics, such as the build-up of ice the detector surfaces (\cref{sec:masterflat}). Using empirical coating models, the estimated contamination rate of $\sim10$\,nm\,month$^{-1}$ is consistent with pre-launch expectations \citep{Schirmer-EP29,DR1-TP026b}. Due to interference, the effective \ac{QE} is variable as a function of time and wavelength \citep[sect.\ 7.7 in][]{DR1-TP026}. For DR1, the differential evolution between \acp{LED} of different wavelengths was still limited, compensated on a per-band basis by relative flux calibration \citepalias[\cref{subsec:throughput_stability} here, and section~ 3.2.2. of][]{Q1-TP003}. In future data releases, these contamination effects will be accounted for using time-dependent corrections of the \ac{QE} curves.

\section{The NIR processing function}
\label{sec:pf}

A detailed description of the \ac{NIR PF} is presented in \citetalias{Q1-TP003} in the context of the Q1 data release. Here, we focus on the main changes and improvements introduced in the current DR1 version with respect to this initial implementation.

\subsection{Common NIR--SIR pre-processing}
\label{sec:preprocessing}

\subsubsection{Bad-pixel identification and masking}
\label{sec:badpix}

In the master \ac{BPM}, `disconnected' pixels {\setcitestyle{notesep={}}\citepalias[see section~ 3.1.3. of][ for a description of bad pixel flavours]{Q1-TP003}} were updated using the \ac{PV} flat-fields \citep[PV-010, ][]{DR1-TP026} taken with {\tt LedId=B}, \texttt{Intensity=60}, and \texttt{PwmDutyCycle=28.7}. `Hot' pixels were updated using 500 PV photo darks \citep[PV-012, ][]{DR1-TP026}.

Permanently `saturated' pixels were missing and are now included in the \ac{BPM}, flagged as {\tt INVALID}.
Saturated pixels are identified across a set of dark current frames.  The method involves a running pair-wise comparison of the frames. For each pair, the pixels that are simultaneously saturated in both frames are identified, and their number is recorded. The pair yielding the minimum number of commonly saturated pixels is then selected, and the corresponding pixel population is taken as an estimate of the permanently saturated pixels. The saturation limit is defined as a threshold close to the A/D conversion limit, accounting for both readout and numerical noise.

Another new category in the \ac{BPM} are pixels forming cross-shaped `flower' patterns, where the central pixel misses the indium interconnect between the HgCdTe detector layer and the readout integrated circuit. The charge in this pixel is largely lost, rendering it dark (`flower centre'), while some signal leaks into its adjacent pixels, which then appear brighter than normal (`flower petals'). These types of defect are known as `open' and `adjacent to open' pixels in the \ac{JWST} data \citep{2014PASP..126..739R}. We identify them using a Laplace filter on a sequence of 30 LED flat-field exposures. A central pixel must be identified in at least two exposures, and is then flagged as \texttt{FLOWER} and \texttt{INVALID}, together with the four adjacent pixels.

The \ac{BPM} is not restricted to scientifically inoperable pixels. To distinguish unusable from informational flavours, the bit 0 -- `INVALID' -- is reserved for the unusable cases. Statistics for the non-transient -- that is static -- bad-pixel types are summarised in \cref{table:badpixdarks}. 
Overall, we identified a total of 239\,878 {\tt INVALID} bad pixels corresponding to approximately 0.4\% of the focal plane. Per detector, we find between 5177 and 31\,700 bad pixels, with a mean of  14\,992.

\subsubsection{Non-linearity correction}
\label{sec:nl}

The non-linearity correction follows the procedure described in Q1 \citepalias{Q1-TP003}, where the per-pixel response is modelled as a 
4th-degree polynomial mapping the measured non-linear flux to a linearised estimate, calibrated against LED flat-field exposures at multiple flux levels. For DR1, the calibration coefficients have been updated using in-flight data from two calibration campaigns: PV-018 (November 2023) and F-008 (February--October 2024), comprising 2136 exposures taken in MACC(4,16,4) mode across six flux levels spanning 8 to 750\,e$^{-}$\,s$^{-1}$ \citep{DR1-TP026}. The highest flux setting (951\,e$^{-}$\,s$^{-1}$) was excluded due to frequent outliers in the final ramp group induced by saturation. The new DR1 non-linearity coefficients are in good agreement with the ground measurements used for Q1, confirming detector stability through launch and commissioning. The validity range of the correction and the treatment of out-of-range pixels remain unchanged; extension to a wider dynamic range is planned for Euclid Data Release 2 (DR2).

\subsubsection{Persistence masking}
\label{sec:persistence}

In Q1 processing \citepalias{Q1-TP003}, the persistence model was derived from ground-characterisation data. This model included only the fast-decay component over short timescales, but the non-linear dependence of the persistence amplitude and decay index (defined below) on the previous illumination intensity was taken into account. Persistence was considered only from spectroscopic to photometric exposures within the same dither (15-minute history), while persistence in spectroscopic exposures from preceding photometric exposures was not allowed for. 

The pipeline was refactored for DR1, and the persistence model updated using in-flight calibration data \citep{DR1-TP038}. The adopted model is based on the decay of persistence following a 60\,ke$^-$ illumination in the PV-016 calibration data set \citep{DR1-TP026}, parameterised as
\begin{equation} 
    I_{\rm P}(S,t)  = \tilde{\alpha} S (t-\tilde{t}_0)^{-\tilde{\beta}}\,,
\label{eq:Ipip}
\end{equation}
where $\tilde{\alpha}$ and $\tilde{\beta}$ denote the initial persistence current amplitude and the decay index, respectively, after a signal $S$, and $\tilde{t}_0$ is the midpoint of the source
exposure. The dependence of $\tilde{\alpha}$ and $\tilde{\beta}$ on the prior illumination intensity is neglected.

\paragraph{Creation of the persistence image}

A persistence image is computed using a dedicated pipeline for every science -- or target -- exposure, by summing the time-integrated power-law decays from the $N$ preceding exposures,
\begin{equation}
    P(S, t, t_{\rm int}) = \sum_{i=1}^{N}\frac{\tilde{\alpha} S_i}{1-\tilde{\beta}}
    \left[ \left(t_i + t_\textrm{int}  \right)^{1-\tilde{\beta}} -  t_i^{1-\tilde{\beta}} \right]\,.
\end{equation}
Here, $t_\textrm{int}$ is the integration time of the target exposure, $S_i$ the signal in the $i$-th preceding exposure, and $t_i$ the time between the midpoints of that preceding exposure and the target exposure.

The working principles are as follows: (i) following a saturation event, a pixel is masked for the full look-back time of 5\,h, as its true signal and therefore persistence are unknown; (ii) persistence contributed by different exposures is approximately additive, a simplification that facilitated pipeline implementation \citep{Kubik2024, DR1-TP038}; (iii) the computation is signal-driven, that is persistence from photo-ionisation and particle ionisation is assumed to behave identical, despite the true amount of charge generated by a particle is masked by the on-board slope fit; (iv) the sky background is ignored, as its persistence signal is assumed to be negligible and mostly removed with background subtraction.

For DR1 processing, the temporal window for persistence computation was set to 5 hours for both photometric and spectroscopic targets. In terms of exposure selection, both photometric and spectroscopic exposures were included when constructing persistence images for spectroscopic targets, while only spectroscopic exposures were considered for photometric targets.

\paragraph{Application of the persistence mask}

The application step uses persistence images produced by the pipeline to identify and flag persistence contamination in science exposures.
For each target and its corresponding persistence image, the persistence signal is evaluated pixel by pixel. Pixels exceeding the relevant threshold are flagged with the \texttt{PERSIST} and \texttt{INVALID} flags in the DQ layer. For photometric targets a threshold of 2\,e$^-$ is used following the current requirements \citep[sect.~5.2 in ][]{DR1-TP038}, while for spectroscopic data, the masking threshold of 30\,e$^-$ was optimised using purity and completeness of spectrometric redshifts \citep{Q1-TP007}. 
The threshold is applied independently of the intensity of the signal in the target exposure for both photometric and spectroscopic targets.
The \texttt{PERMODFAIL} flag is raised for pixels where the persistence estimate may be unreliable, for example, if a pixel in the history exposures is saturated or affected by cosmic rays and energetic particle hits, or if the persistence model coefficients are undefined. The science exposure header is updated with the threshold values used, along with summary statistics per detector, including the number of masked, corrected, failed, and invalid pixels.

For photometric targets, an Hough transform \citep[HT;][]{Hough1962} line detection is applied to the target exposure to enhance masking of persistence streaks. Pixels belonging to detected linear features are flagged with \texttt{PERSIST} in the DQ layer if they have not already been flagged via thresholding of the persistence image.
For the implementation on \Euclid images, a threshold is first applied on the image, and the positions of surviving pixels are used for the calculation. To avoid flagging diffraction spikes, which remain attached to their stars after thresholding, connected components larger than 10\,000 pixels are removed prior to the HT, while a minimum line length of 200 pixels excludes the shorter spikes of fainter stars. The probabilistic version of the HT has been used here, specifically \texttt{scikit-image}’s \texttt{probabilistic\_hough\_line} function, since this is the version of the method that is capable of returning the end points of the detected lines in addition to their location and orientation, and is moreover much less computationally expensive. When applied within the \ac{NIR PF} as a step following the power-law masking, the HT offers a geometric approach to search for any residual linear structures in the images that remained undetected. This contributes an additional 2\% of the number of pixels flagged by the power-law model, which is an indication that the latter is successful in masking the vast majority of persistence traces in the images, leaving few remainders for the HT to detect.


\begin{figure}[t]
    \centering
    \includegraphics[width=\linewidth]{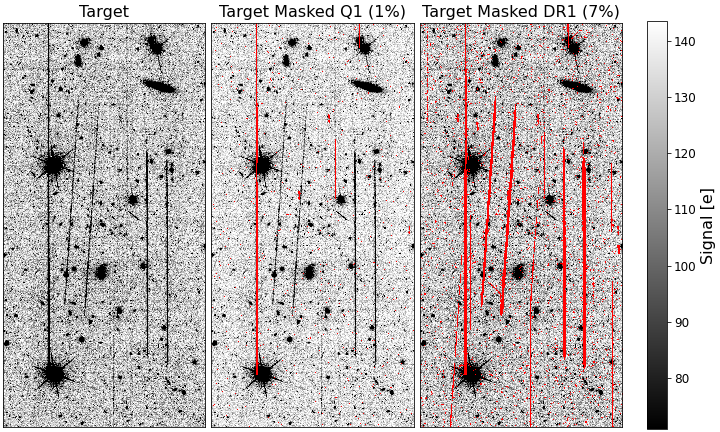}
    \caption{Comparison of persistence masking in Q1 and DR1 in a 500\,$\times$\,1000-pixel region. Masked pixels are shown in red. Persistence from some spectra is unmasked in Q1 but masked in DR1 due to the longer history considered when computing the persistence image. The fraction of masked pixels is indicated in parentheses above each panel.}
    \label{fig:masking_dr1_q1}
\end{figure}

\paragraph{Persistence masking: Q1 versus DR1}

\begin{figure}[h!]
    \centering
    \includegraphics[width=\linewidth]{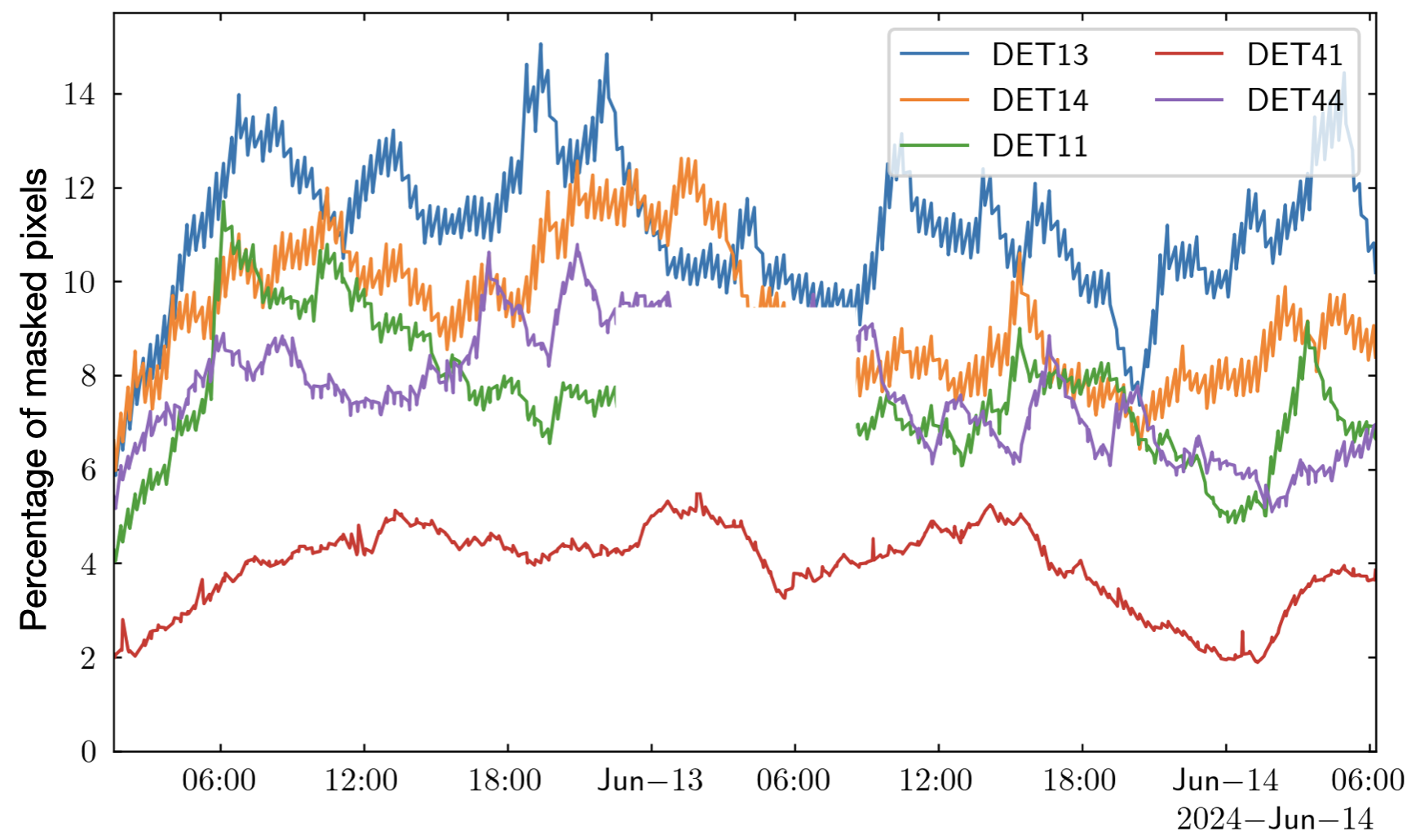}
    \caption{Fraction of masked pixels in DR1 processing for the five detectors with the highest persistence in SkyPatch\,38 during a period of high solar activity. For some detectors, the masking fraction exceeds 10\% due to the high flux of energetic particles and limitations of the persistence model.}
    \label{fig:masking_fraction_dr1}
\end{figure}

\Cref{fig:masking_dr1_q1} highlights the improvement in persistence masking in DR1 with respect to Q1 processing. Persistence from some spectra is unmasked in Q1, but correctly masked in DR1, due to the longer illumination history considered in the pipeline.
However, DR1 processing sometimes masks too many pixels because the masking parameters are very conservative \citep{DR1-TP038}. \Cref{fig:masking_fraction_dr1} shows the fraction of pixels masked in DR1 processing for SkyPatch\,38, consisting of 43 observations in the EDF-North \citep{Scaramella-EP1} acquired during a period of high solar activity. Several detectors exhibit masking above 10\%. This is partly caused by the enhanced flux of energetic particles during high solar activity, and partly by limitations of the persistence model, which does not fully account for dependencies on signal intensity, nor the different behaviour of persistence generated by photo- and particle ionisation.

While the model presented in \citet{DR1-TP038} and adopted in this work successfully identifies and masks pixels affected by persistence, accurately predicting the associated excess flux remains challenging. Consequently, a further increase in the number of usable pixels may be achieved by applying persistence corrections rather than relying on masking alone. Several alternative correction approaches are currently being investigated to improve the accuracy of the correction and assess their suitability for potential inclusion in the Euclid DR2 processing pipeline and subsequent data releases. A comprehensive comparison of these approaches will be presented in Awad et al. (in prep.).

\subsection{Astrometric and photometric calibration of NISP images}
\label{sec:astrophoto}

\subsubsection{Pixel-response non-uniformity (PRNU) correction}
\label{sec:masterflat}

The Euclid DR1 processing corrects PRNU in all three bands using filter-dependent master flats derived from \ac{LED} exposures, following \citetalias{Q1-TP003}. Each of the five \acp{LED}, designated A to E, in the NI-CU produces a near-monochromatic \ac{SED} and illuminates the \ac{FPA} directly at a small off-axis angle, without passing through a filter \citep{EuclidSkyNISPCU}. The \acp{LED} B, C, and D provide the exposures used to derive the \YE, \JE, and \HE master flats, respectively. For each \ac{LED}, we calibrate 30 exposures using the pre-processing steps in \cref{sec:preprocessing} and combine them into a master flat. Section~3.2.1 and appendix~A of \citetalias{Q1-TP003} describe the conversion from calibrated \ac{LED} flats to filter-dependent master flats using the \ac{QE} map and filter transmission curve.

We used one set of filter-dependent master flats throughout DR1 processing. To assess temporal changes, we examined 24 \HE master flats constructed from LED-D exposures between March 2024 and April 2026. This monitoring sequence covers part of the DR1 observing period and extends beyond it. 

We denote the master flat for detector $d$ at epoch $t$ and pixel $\mathbf{x}$ by $\mathcal{F}_d(\mathbf{x},t)$, and use the same set of valid pixels, $\Omega_d$, at every epoch. We isolate the small-scale component $S_d(\mathbf{x},t)$ as
\begin{equation}
S_d(\mathbf{x},t)
=
\frac{\mathcal{F}_d(\mathbf{x},t)}
{G_\sigma[\mathcal{F}_d(\mathbf{x},t)]}\,,
\label{eq:prnu-small-scale}
\end{equation}
where $G_\sigma$ is a mask-aware Gaussian smoothing operator with standard deviation $\sigma=64$ pixels. This procedure effectively suppresses large-scale variations while preserving local pixel-to-pixel fluctuations. For a fixed mask, a detector-wide multiplicative normalisation cancels between the numerator and denominator. Mask construction and smoothing checks are described in \cref{app:prnu-methods}.

We characterise spatial variability using
\begin{equation}
\sigma_{68}(S)
=\frac{P_{84}(S)-P_{16}(S)}{2}\,,
\label{eq:prnu-sigma68}
\end{equation}
where $P_p$ is the $p$th percentile over valid pixels. The median static PRNU amplitude, $\sigma_{68}(S_d)$, across the 16 detectors changes from 1.269\% at the first epoch to 1.272\% at the final epoch. The change is 0.0029 percentage points, or 0.23\% of the initial amplitude. This statistic measures the spatial structure within each master flat; its stability does not imply that every pixel remains unchanged.

We measure temporal variability using the logarithmic residual
\begin{equation}
\Delta\ln S_d(\mathbf{x};t_i,t_j)
=\ln\!\left[\frac{S_d(\mathbf{x},t_j)}{S_d(\mathbf{x},t_i)}\right]\,.
\label{eq:prnu-log-residual}
\end{equation}
For small changes, this residual approximates the fractional change in the small-scale component; percentages below denote $100\,\Delta\ln S_d$. The early reference, $S_{d,\mathrm{early}}$, is the pixel-wise median of the first three epochs. We compare subsequent epochs with this reference and consecutive epochs with one another.

To test whether the residuals grow with temporal separation without selecting a fixed reference, we calculate a robust temporal structure function (SF) as
\begin{equation}
\mathrm{SF}(\tau)
=\frac{1}{\sqrt{2}}\,
\operatorname{median}_{(d,i,j)\in\mathcal{P}_\tau}
\left[\sigma_{68}\!\left(\Delta\ln S_d\right)\right]\,,
\label{eq:prnu-structure-function}
\end{equation}
where $\mathcal{P}_\tau$ contains all detectors and distinct epoch pairs with $i<j$ and $t_j-t_i$ in lag bin $\tau$. We first calculate $\sigma_{68}$ over the valid pixels of each detector and then take the median over detector--pair measurements. The $1/\sqrt{2}$ factor gives a single-epoch-equivalent amplitude for independent, approximately Gaussian fluctuations of comparable amplitude.

\begin{figure}
\centering
\includegraphics[width=0.98\linewidth]{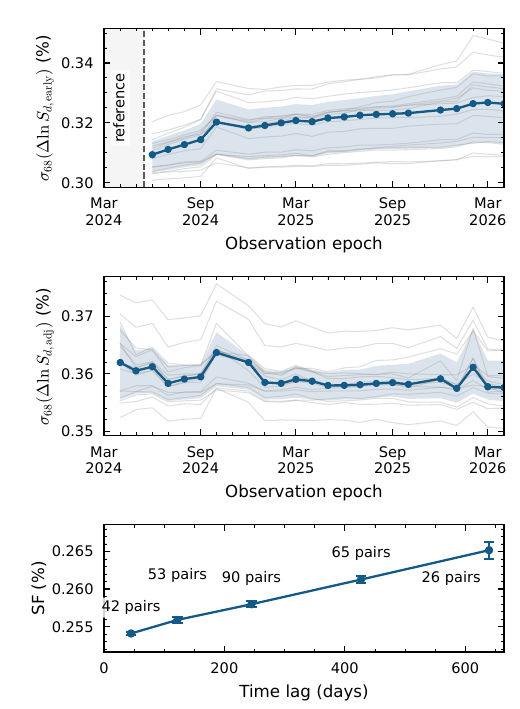}
\caption{Small-scale flat-field stability. \emph{Top:} The scatter relative to the pixel-wise median of the first three epochs. The shaded interval marks those reference epochs; only subsequent measurements are plotted. \emph{Middle:} Adjacent-epoch repeatability of the small-scale flat-field. Thin grey curves represent individual detectors, blue curves show the median across detectors, and shaded bands span the 16th--84th detector percentiles. \emph{Bottom:} The structure function defined in \cref{eq:prnu-structure-function}. Error bars are leave-one-epoch-out jackknife standard errors; labels give the number of distinct epoch pairs in each lag bin. All ordinates are percentages.}
\label{fig:flat-stability-temporal-v2}
\end{figure}

At the final epoch, the median reference-relative scatter across detectors is 0.326\%, with a range of 0.309--0.347\% (\cref{fig:flat-stability-temporal-v2}a). Across adjacent-epoch comparisons of all detectors, the median scatter is 0.359\%, with 16th--84th percentiles of 0.356--0.364\%; no focal-plane-wide discontinuity is seen (\cref{fig:flat-stability-temporal-v2}b). The structure function rises from $0.2541\% \pm0.0002\%$ below 90 days to $0.2652\% \pm0.0012\%$ at 550--800 days (\cref{fig:flat-stability-temporal-v2}c). These uncertainties are leave-one-epoch-out jackknife standard errors. The rise remains positive when any epoch is omitted. The short-lag value is consistent with $0.359\%/\sqrt{2}$; the higher long-lag value indicates a time-dependent contribution to the differences between delivered master flats, although without identifying its origin.

We also compare pixel-wise median stacks of the final three and first three epochs. Pooling all valid focal-plane pixels, this late-to-early residual has $\sigma_{68}=0.264\%$, with scatters of 0.246--0.291\% across all detectors. Its absolute residuals have $P_{95}(|\Delta\ln S|)=0.534\%$ and $P_{99}(|\Delta\ln S|)=0.734\%$ (\cref{fig:flat-stability-fpa-v2}). The late-to-early residual map in \cref{fig:flat-stability-fpa-v2} shows detector-boundary structures with positive residuals, compact positive regions, and several extended features. The selected compact positive regions have a median area-equivalent diameter of 5.4\,pixels and a median local contrast of 0.22\%. Their selection and the measurements of other residual structures are described in \cref{app:prnu-features}.

The residuals include statistical uncertainty in master-flat construction and possible contributions from the calibration source, persistence, and processing. We have not subtracted a measurement-noise term. The reported scatters therefore measure repeatability of the delivered master flats, not intrinsic detector evolution. Changing the smoothing scale to 32 or 128 pixels alters the final reference-relative scatter by less than 1\% relative to the nominal result. The sub-percent temporal scatter is consistent with a fixed small-scale PRNU correction for DR1.

\subsubsection{Background estimation}
\label{sec:background}

For the background estimation step of the DR1 processing, the {\tt GNUastro} tool {\tt NoiseChisel} v0.16 \citep{Akhlaghi2015} has been used, replacing {\tt SourceExtractor} \citep{Bertin1996}, which was initially part of the NIR preparation of the Q1 data release \citepalias{Q1-TP003}.  

{\tt NoiseChisel} estimates the background by first applying a quantile threshold (set to 0.3 or 30th quantile) to create a binary map in which pixels above the threshold are marked as foreground. The thresholded image then undergoes a series of morphological operations, including erosion, opening to remove isolated clumps, and dilation to recover faint source signal. This produces a final source mask. After the pixels from this mask are excluded, the image is tessellated into tiles (set to a size of 10\,$\times$\,10 pixels), and the background in each tile is computed from the remaining sky pixels. The background estimates in each tile are then interpolated across the image using a Manhattan-distance-based nearest-neighbour scheme\footnote{See the \href{https://www.gnu.org/software/gnuastro/manual/html_node/Interpolation.html}{{\tt GNUastro} documentation on interpolation}.}, and smoothed with a flat kernel set to a width of 3 pixels. This configuration was found to provide the best compromise between preserving the surface brightness of faint, extended sources, and minimising residuals associated with bright objects. Additionally, and similarly to what was done for the Q1 NIR processing, the estimated background is not directly subtracted from the corresponding scientific image, but is provided as a companion file.  

\begin{figure*}[h!]
    \centering
    \includegraphics[width=\textwidth]{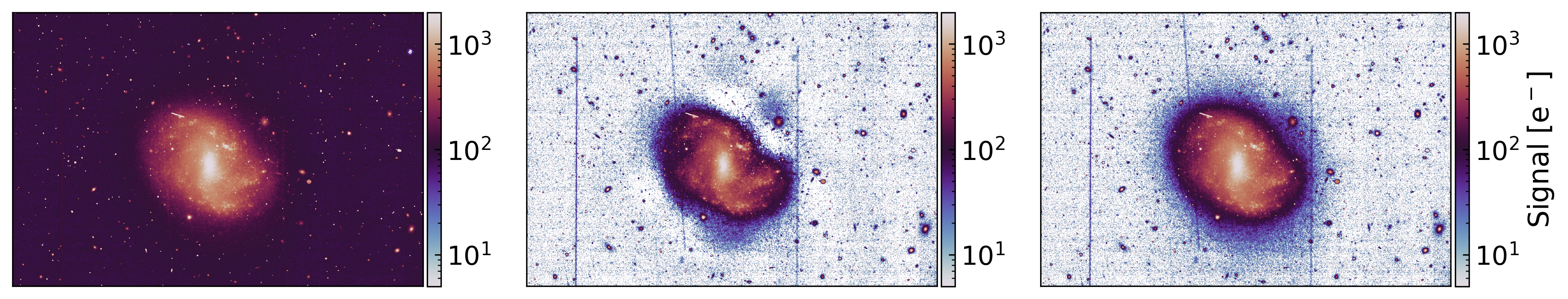}
    \caption{{\it Left:} A 750\,$\times$\,500-pixel section of a \Euclid NIR image in the \JE band containing the barred spiral galaxy NGC~1483. {\it Middle:} The same image after background subtraction using the background estimated by {\tt Source-Extractor}. {\it Right:} The background-subtracted image using the sky background estimated by {\tt NoiseChisel}, showing the improved preservation of faint light surrounding the galaxy.}
    \label{fig:BkgSub}    
\end{figure*}

A demonstration of the quality of the background subtraction is shown in \cref{fig:BkgSub}, where the subtraction is performed using the {\tt SourceExtractor} and {\tt NoiseChisel} estimated backgrounds in the middle and right panels respectively. The improvement is visible in the preservation of faint light and low surface-brightness features surrounding the galaxy. Furthermore, the subtraction of the background reveals a faint striping pattern within the images, which is attributed to $1/f$ noise in the detector readout. This effect will be treated and integrated within the background estimation scheme for the forthcoming release (DR2), enabling a more homogeneous image background while maximising the preservation of source brightness.

\subsubsection{PSF modelling from data}
\label{sec:psf}

For Q1, the \ac{NIR PF} derived the \ac{PSF} models on-the-fly during the processing of each individual image \citepalias{Q1-TP003}. While this approach allowed the \ac{PSF} to be estimated directly from the data being processed, the limited number of suitable sources available in a single exposure constrained the accuracy that could be achieved, particularly in reproducing the \ac{PSF} wings. In some cases, this also resulted in an unstable solution at the level of individual detectors. For DR1, the \ac{PSF}-modelling strategy has been revised \citep{DR1-TP005}, and the \ac{PSF} for individual images is now derived in advance using a dedicated set of observations of the self-calibration field, comprising 76 dithers \citep{DR1-TP026}. The larger and more homogeneous data set provides a significantly improved constraint on the PSF, resulting in more accurate models and enhanced stability. The resulting \ac{PSF} models are stored in the \ac{EAS} and retrieved at run time by the NIR PF when processing the science images.

In the following, we provide a brief description of the new \ac{PSF}-modelling approach and present its main results, while referring the reader to the dedicated paper \citep{DR1-TP005} for a detailed description of the methodology and its validation, including the analysis of possible \ac{PSF} spatial, colour, and temporal variations.

{\tt SourceExtractor} \citep{Bertin1996} is used to detect and select suitable point-like sources from individual images, while {\tt PSFEx} \citep{Bertin2011} is employed to derive the corresponding \ac{PSF} models. The Q1 data were used to support the development of the \ac{PSF}-modelling and catalogue-extraction modules of the \ac{NIR PF}, as well as to refine the \ac{PSF} models in preparation for DR1.

The procedure of fitting the \ac{PSF} profile, compared to other methods used for flux estimation, such as aperture photometry, indicates that the \ac{PSF} models under production for DR1 are better than those derived from Q1 data. In \citet{DR1-TP005}, a comparison was made between {\tt MAG\_PSF} and {\tt MAG\_APER}, both magnitudes provided by {\tt SourceExtractor}, as a function of {\tt MAG\_APER} for magnitudes in the range 17--21. When going from Q1 to the DR1 data and using an improved version of the \ac{PSF} and catalogue extraction modules (\cref{sec:catalog}), the results, summarised in \cref{tab:psf_comp}, show a reduction in photometric residuals by a factor 2 to 5 for the three filters, and improved consistency between the two types of photometry across the magnitude range.

\begin{table}
\centering
\caption{Comparison of PSF and aperture photometry for Q1 and DR1.}
\label{tab:psf_comp}
\setlength{\tabcolsep}{5pt}
\begin{tabular}{lcccc}
\hline
\hline
\noalign{\vskip 1pt}
Band & \multicolumn{2}{c}{Slope} & \multicolumn{2}{c}{RMS [mag]}  \\
\hline
\noalign{\vskip 1pt}
& Q1 & DR1 & Q1 & DR1 \\
\hline
\noalign{\vskip 1pt}
\YE & $0.0192\pm0.0028$ & $0.0040\pm0.0008$ & $0.074$ & $0.017$ \\
\JE & $0.0045\pm0.0010$ & $0.0017\pm0.0005$ & 0.026 & 0.011 \\
\HE & $0.0024\pm0.0006$ & $-0.0019\pm0.0004$ & 0.018 & 0.010 \\
\hline
\end{tabular}
\end{table}

The \ac{PSF} models based on the stacked frames were not among the data products delivered in Q1, and no particular changes have been included for DR1. However, the same improvements implemented in the catalogue-extraction module for single frames can be adopted for stacked frames. By performing a similar comparison as described above, preliminary results do not suggest significant improvements in the distributions of the residuals. Tuning of the {\tt PSFEx} parameters is expected to improve results further, and is planned for future releases.

\subsubsection{Astrometric calibration}
\label{sec:astrometry}

A detailed description and analysis of the results for the DR1 astrometric calibration is presented in \citet{DR1-TP004}; here, we provide only a short summary. As in Q1, the astrometric calibration in DR1 is based on {\tt SCAMP} \citep{Bertin2006}.
The astrometric solution is obtained by using the \ac{PSF} centres of sources detected by {\tt SourceExtractor}, along with reference positions from the {\it Gaia} DR3 catalogue, which have been adjusted for proper motion to match the dither epoch \citep{GaiaDR3}. A third-order polynomial is fitted to each detector, minimising the sum of squared positional differences between matched sources both from the {\it Gaia} reference stars, and those detected in overlapping exposures. The robustness of the matching process is enhanced by employing an initial distortion (pre-solution), which is based on observations from the self-calibration field. 
In DR1, only two minor changes were made compared to Q1. (i) Before the astrometric calibration, an additional step was introduced to identify exposures that pose problems for accurate astrometric calibration, such as failures in the Fine Guidance Sensor that result in elongated source images or energetic particle showers that produce an extremely high number of spurious detections. (ii) To transfer the pre-solution to each dither, it is essential to consider the dither's rotation on the sky. This is achieved by converting the pre-solution into the simple image polynomial (SIP) formalism \citep{Shupe2005}, and then back to PV adjusting the CD matrix to account for the instrument rotation. In Q1, this transformation was achieved through polynomial fitting, while in DR1, an analytic conversion is employed.

\subsubsection{Catalogue production}
\label{sec:catalog}

In the DR1 version of the catalogue extraction step, we implemented a multi-aperture photometry approach that provides flux measurements using a set of concentric apertures with radii of 0.6, 0.9, 1.5, 2.25, and 3\arcsecond\ (corresponding to diameters of 4, 6, 10, 15, and 20\,pixels). Multi-aperture photometry is important for accurately characterising sources with different spatial extents. Small apertures provide robust flux measurements for compact sources while minimising background contamination, whereas larger apertures capture a larger fraction of the flux from extended sources and provide a more reliable estimate of their total integrated brightness. The resulting flux growth curves also allow us to quantify flux losses in small apertures, estimate the encircled energy fraction, and assess the source spatial profile relative to the \ac{PSF}. These measurements are therefore relevant both for photometric calibration and for the characterisation of source morphology.

\begin{figure*}
\centering
\begin{minipage}[t]{\textwidth}
    \centering
    \includegraphics[width=0.49\linewidth]{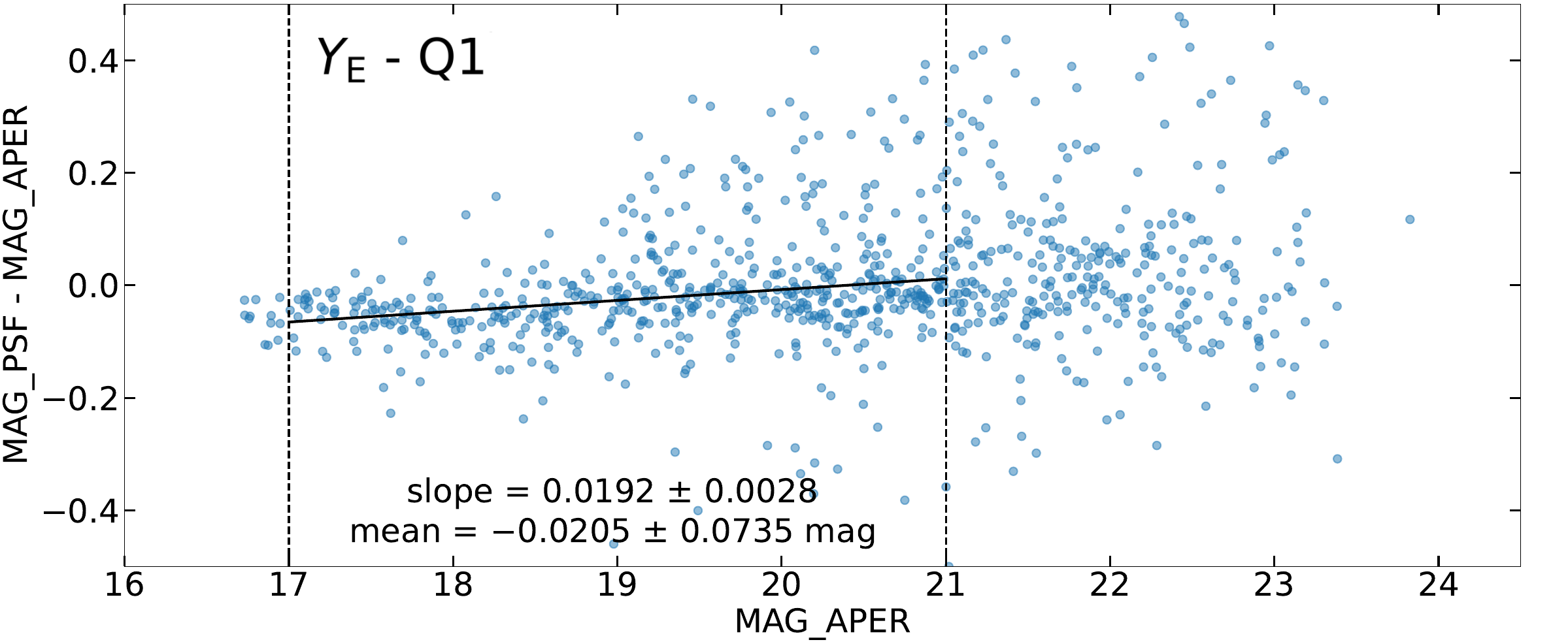}
    \includegraphics[width=0.49\linewidth]{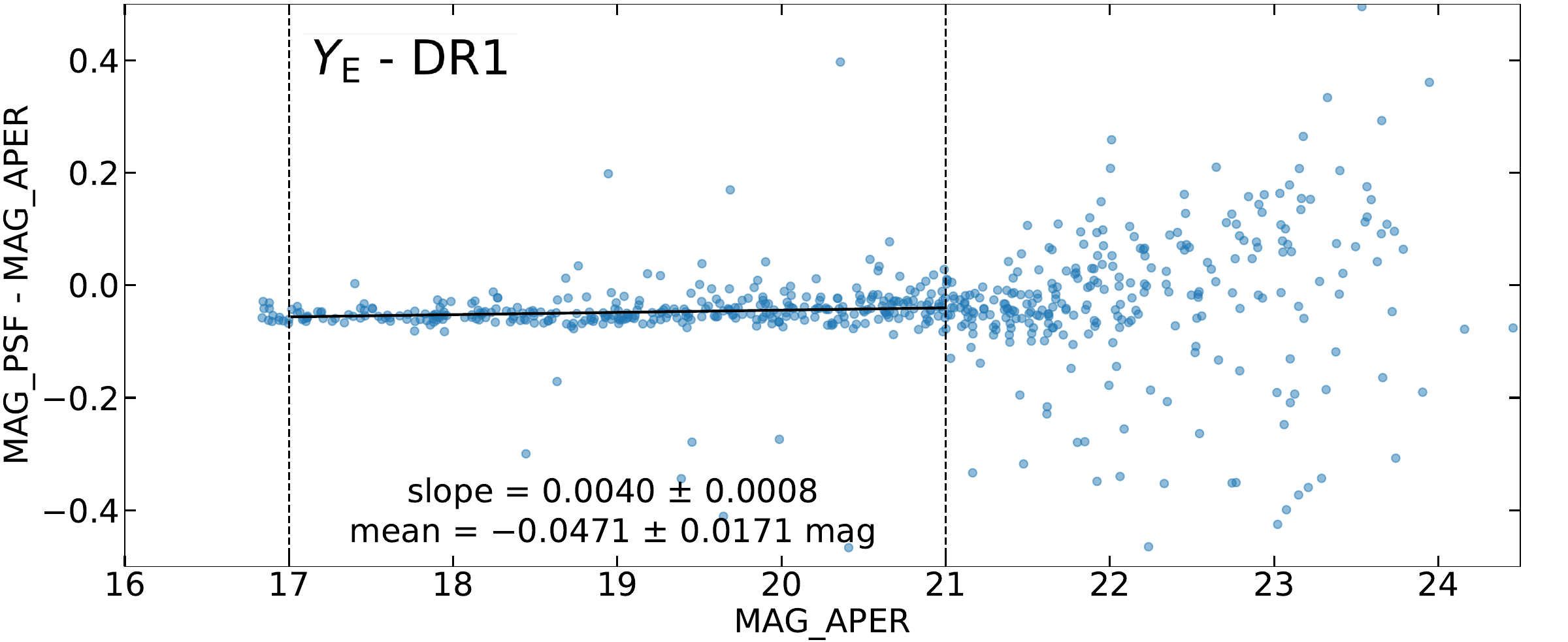}
    \end{minipage}
    \caption{\ac{NIR PF} Catalogue Extraction: Comparison of PSF-fitting and aperture photometry ({\tt MAG\_PSF} $-$ {\tt MAG\_APER}) for the \YE band across Q1 (left panel) and DR1 (right panel) versions of code and \ac{PSF} models.}
    \label{fig:phot_improv}
\end{figure*}

We also tuned task parameters, resulting in a more effective rejection of spurious detections and improved \ac{PSF}-fitting photometry. The Q1 code version was optimised using simulations and data from \Euclid's \ac{PV} phase \citep{DR1-TP026}. While fixed-aperture photometry provided satisfactory results, the \ac{PSF}-fitting photometry showed systematic biases. One of the main issues was the use of a high initial detection threshold, which led to an underestimation of the source extent in the source map. {\tt SourceExtractor} uses this source map to define the region over which the \ac{PSF} model is fitted; consequently, underestimated source sizes resulted in biased flux measurements \citepalias[see sect.~3.2.6 in][]{Q1-TP003}. In the DR1 version, the initial detection threshold was lowered from 5 to 2, allowing the algorithm to converge adaptively towards a more appropriate threshold and hence improving the \ac{PSF}-fitting results. This improvement is illustrated by the comparison of the two plots in \cref{fig:phot_improv}, which shows the results obtained in the \YE band. A detailed description of the development and performance of the \ac{PSF} and catalogue extraction modules of the NIR pipeline, including results for all three NISP bands, is presented in \citet{DR1-TP005}.

Two additional selection criteria, {\tt FLUX\_RADIUS}\,$>$\,0.65 and {\tt FWHM\_IMAGE}\,$>$\,0, were introduced to improve the rejection of spurious detections by exploiting the signal-to-noise ratio and measured size of the detected sources. In particular, since the NIR source catalogues are also used to generate several calibration products, we adopted a conservative selection criterion requiring the {\tt SourceExtractor} quality flag to be null. This criterion removes potentially problematic, yet valid, detections from the source catalogue, including blended sources.

\subsection{Resampling and stacking of NISP images}
\label{sec:resandstack}

\begin{figure}[h]
\centering
\resizebox{\hsize}{!}{\includegraphics{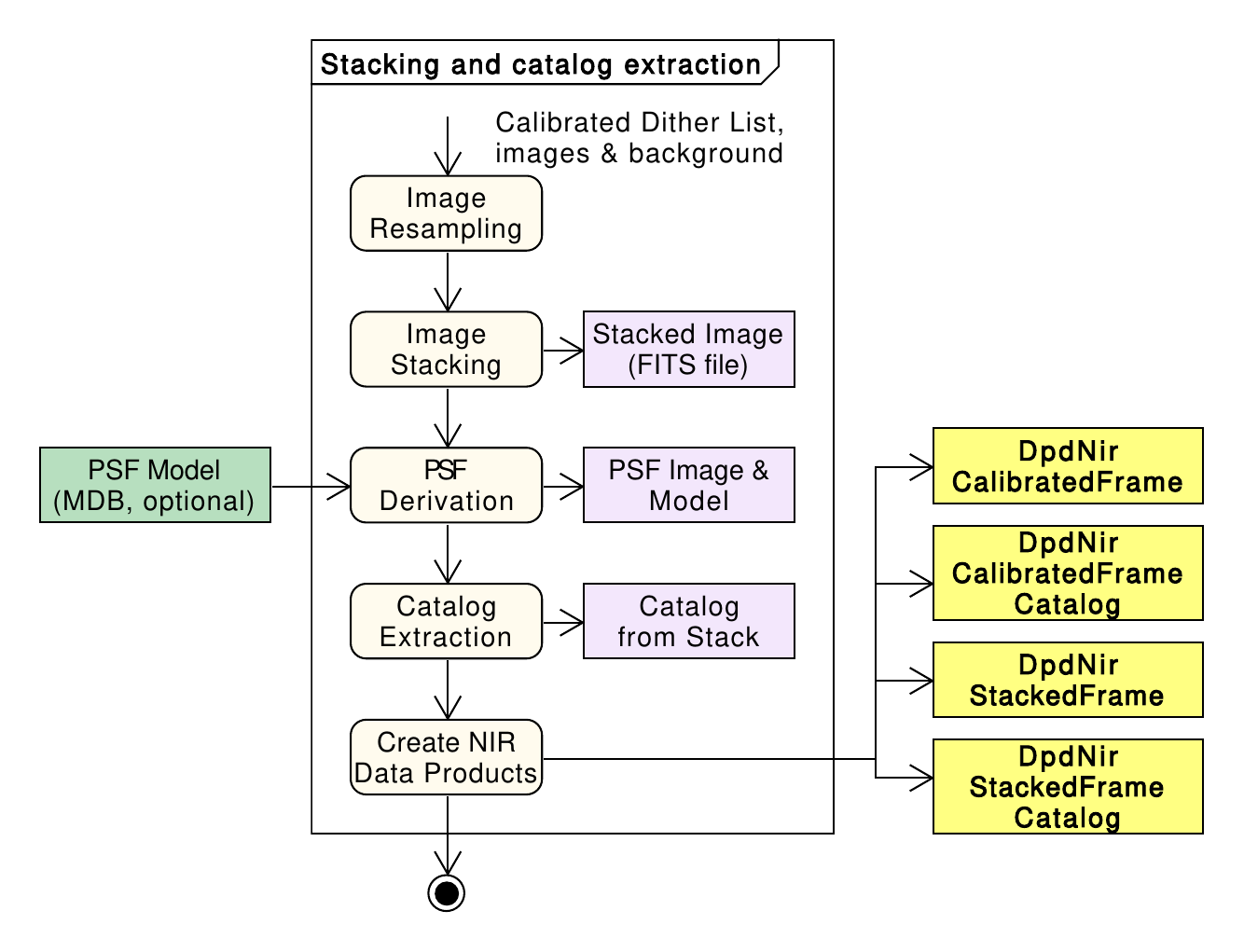}}
\caption{\ac{NIR PF} resampling and stacking. The last \ac{PE} is dedicated to the creation of the final \acp{DP}, represented as yellow boxes, that are stored into the \ac{EAS}.}
\label{fig:res&stack}
\end{figure}

In this section, we describe the \acp{PE} in the \ac{NIR PF} that are designed to combine single-epoch images of the same field and filter into a co-added frame. The corresponding activity diagram is shown in \cref{fig:res&stack}. This represents a new addition to the pipeline, since these processing steps and the associated \acp{DP} were not included in the Q1 release.

The last \ac{PE} is dedicated to the creation of the output \acp{DP}, represented as yellow boxes in \cref{fig:res&stack}.
In addition to the calibrated single-epoch image, {\tt DpdNirCalibratedFrame}, and its corresponding source catalogue, {\tt DpdNirCalibratedFrameCatalog}, already released for Q1 \citepalias{Q1-TP003}, the \ac{NIR PF} in DR1 produces two new \acp{DP}: the stacked image, {\tt DpdNirStackedFrame}; and its corresponding source catalogue, {\tt DpdNirStackedFrameCatalog}. 
Detailed information on the definition of the NIR \acp{DP} for DR1 are reported in \cref{app:dataproducts}.

\subsubsection{Resampling}
\label{sec:resampling}

The main goal of the \texttt{NIR\_resampling} module is to realign, rotate, and resample onto a common grid the 16 detector images from the dither exposures acquired in the same field and with the same NISP-P filter. The astrometrically and photometrically calibrated images are first background-subtracted using the background estimates described in \cref{sec:background}, and  then processed using {\tt SWarp} \citep{Bertinetal2002}, which projects each detector image onto a common grid. {\tt SWarp} uses an inverse-mapping technique: the output frame is scanned pixel by pixel, and the position of each output pixel in the input frame is determined through the inverse geometrical transformation. Since the resulting coordinates generally do not coincide with the input pixel grid, the corresponding pixel value is estimated by interpolating the values of neighbouring input pixels. This approach ensures that every pixel in the output frame is assigned a value, avoiding the gaps that may arise with forward-mapping techniques.

We tested nearest-neighbour, bilinear, and Lanczos-2, -3, and -4 interpolation functions \citep{SwarpUserManual} by comparing the magnitude and centroid differences of extracted sources before and after resampling. Lanczos-2 provides a good compromise, yielding a mean signed difference of less than 0.005\,mag in magnitude and an RMS of 0\farcs015 in both RA and Dec, while introducing fewer artefacts around image discontinuities. For comparison, bilinear interpolation yielded a mean signed difference of 0.010\,mag and RMS residuals of 0\farcs020 in both RA and Dec. Lanczos-3 performed slightly better in terms of accuracy, with a mean signed difference of 0.003\,mag and RMS residuals of 0\farcs010, but its larger 6\,$\times$\,6-pixel kernel significantly increased the number of flagged pixels.
Resampling is performed with an oversampling factor of 5, with fluxes rescaled according to the local pixel-scale ratio, and the inverse variance maps of the input dithers are used as weights.
An additional flux scaling factor, accounting for the exposure time and the inverse sensitivity associated with each NISP-P filter (\cref{sec:abs_phot}), is computed during resampling and applied when stacking the individual frames to convert the fluxes to the adopted AB magnitude system.

\begin{table}
\caption{Parameters in the {\tt SWarp}  configuration file \citep{SwarpUserManual} used by the resampling module.}
\centering
\smallskip
\label{table:swarp_params}
\smallskip
{\small
\begin{tabular}{ll}
\hline\hline\noalign{\vskip 2pt}
{\tt SWarp} Parameter & Value \\
\noalign{\vskip 2pt}\hline\noalign{\vskip 2pt}
{\tt FSCALE\_KEYWORD} & set at runtime  \\
{\tt FSCALASTRO\_TYPE} &  {\tt VARIABLE} \\
{\tt OVERSAMPLING} & 5 \\
{\tt RESAMPLING\_TYPE} & {\tt LANCZOS2} \\
{\tt WEIGHT\_TYPE} & {\tt MAP\_VARIANCE} \\
{\tt COMBINE\_TYPE} & {\tt WEIGHTED} \\
\hline
\end{tabular}}
\end{table}
The resampled images from each detector and dither exposure are combined into a \ac{MEF} file preserving the three-layer structure -- image (SCI), RMS, and \ac{DQ} -- of the input dithers \citepalias{Q1-TP003}. Flags from the \ac{DQ} layer are also propagated. Since this functionality is not available in the official version of {\tt SWarp}, we use a dedicated C++ version of the code. Overlapping adjacent pixels are combined using the OR operator: if the flags are identical, the original flag is retained; if they differ, the resulting flag is the bit-wise sum of the two flag values. The resampled variance layer is obtained as the inverse of the resampled weights.

\subsubsection{Stacking}
\label{sec:stacking}

The astrometrically and photometrically calibrated dither exposures for a given field and filter, previously resampled onto a common grid, are combined into a single stacked image using {\tt SWarp}.
As for the resampling, the stacking procedure is also controlled by a set of parameters defined in the configuration file. The {\tt COMBINE\_TYPE} parameter specifies how the input images are combined. We tested several options and found that the `weighted' and `median' methods provide the best results by visual inspection. The parameter is currently set to `weighted', while {\tt WEIGHT\_TYPE} is set to `map\_weight'. With this configuration, for each output pixel, {\tt SWarp} combines the corresponding values from all input images, weighting them according to the resampled weight maps.

The current version of the \texttt{NIR\_stacking} module also supports the propagation of flags stored in the \ac{DQ} layer. This functionality is implemented using the C++ version of {\tt SWarp}. Flags from different dithers are combined using the OR operator as for the resampling step. Flag propagation is controlled by a boolean input port, which is currently set to `false', but can be enabled if needed.

The combined frames are stored in a three-layer (SCI, RMS, and \ac{DQ}) \ac{MEF} file. The co-added RMS layer is computed as the square root of the co-added variance, which is obtained as the inverse of the co-added weights. For zero weights, the variance is set to $10^{16}$, which is significantly higher than the variance of valid pixels, ensuring that they are effectively excluded from the weighted stacking without introducing numerical issues. Since flag propagation is currently disabled, the co-added DQ layer is derived from the co-added weights by setting pixels with zero weight to 1 and all other pixels to 0. Finally, the headers of the \ac{MEF} files are populated with the keywords required by the Euclid Common Data Model.

\section{Results for Euclid Data Release 1}
\label{sec:imageaccuracy}

\subsection{Accuracy of the astrometric calibration}
\label{subsec:astrometry_res}
The astrometric performance and the limitations identified during the DR1 processing are described in detail in \citet{DR1-TP004}. The achieved accuracy is well within the requirement of 30\,mas. When all detections in a given dither are considered, the offsets with respect to {\it Gaia} are within 2\,mas; the average RMS is 10\,mas and remains below 20\,mas for 99\% of the dithers in all bands. Similar values are obtained when considering overlaps across dithers in each band.

During DR1 processing, the more conservative approach adopted to flag pixels potentially affected by persistence sometimes resulted in a notable decrease in the number of sources with matches in the {\it Gaia} catalogue, from approximately 30 per detector to just a few. This effect was particularly noticeable for detectors DET13 and DET14 {\setcitestyle{notesep={}}\citep[see ][ for the definition of the detector's slot]{EuclidSkyNISP}}. Consequently, this could lead to a local reduction in astrometric accuracy to above 50\,mas for those two detectors \citep[see also][]{DR1-TP026}. While the issue was resolved, the corresponding correction could not be implemented in DR1 and will instead be applied in DR2. A comparative analysis with the VIS catalogues indicates that this issue impacts approximately 3\% and 0.3\% of the total DR1 area, with RMS thresholds of 30\,mas and 60\,mas, respectively.
Regions affected by this issue can be identified by VMPZ-ID {\tt InfoMaps} \citep{DR1-TP028}, which are based on the RMS maps also described in \citet{DR1-TP004}.

\subsection{Stability of the NISP throughput}
\label{subsec:throughput_stability}

\begin{figure*}[ht!]
    \centering
    \begin{minipage}[t]{0.98\textwidth}
    \includegraphics[width=0.47\linewidth]{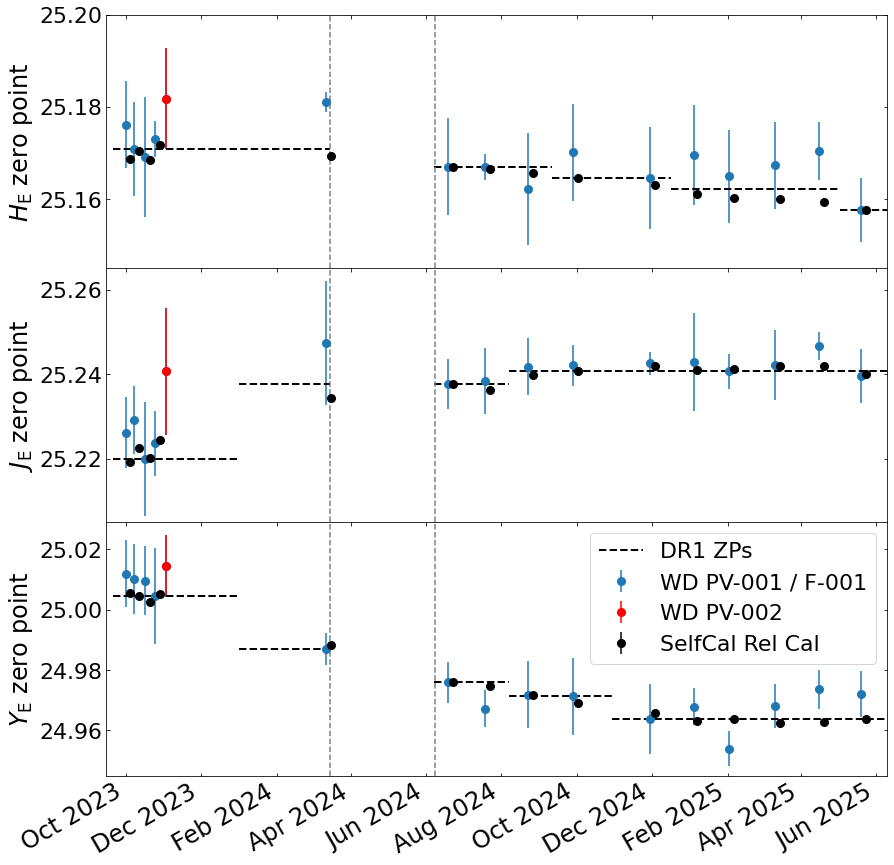}
    \includegraphics[width=0.47\linewidth]{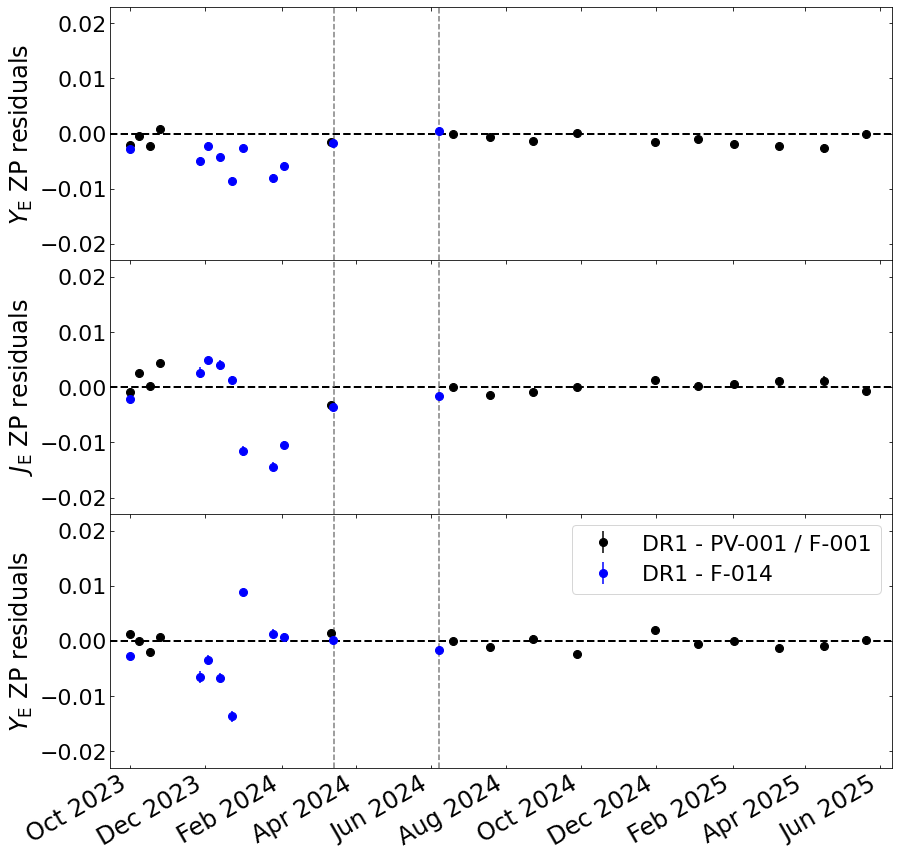}
    \end{minipage}
    \caption{{\it Left:} Monitoring of the NISP zero-point evolution using both the WD calibrator (blue and red points) and the combination of all sources (black points) from the monthly self-calibration field visits. Black dashed lines represent the zero points adopted for DR1. {\it Right:} Residuals for the NISP zero points throughout the DR1 period. Results are compatible with the 1.5\% requirement on relative photometric calibration.}
    \label{fig:WD_ZP_selfcal}
\end{figure*}

Similarly to Q1, we take advantage of the monthly visits to the self-calibration field to monitor the stability of the \ac{NISP} throughput and adjust the photometric zero point to maintain the relative calibration within the required 1.5\% accuracy. In addition to monitoring the \ac{WD} primary calibrator WDJ175318.65+644502 (\cref{sec:abs_phot}), we follow the procedure described in section~5.2 of \citetalias{Q1-TP003} to compute magnitude differences for each visit, adopting the self-calibration field observation from June 2024 as the reference epoch, which corresponds to the observation immediately following the second decontamination procedure \citep{DR1-TP026b}.
The results shown in \cref{fig:WD_ZP_selfcal} indicate variations of a few percent in the \ac{NISP} throughput in all three \ac{NIR} filters. These variations are consistent with the formation of ice on the detectors \citep{DR1-TP026}. It is worth noting that, during each self-calibration visit, the \ac{WD} is observed on only a limited number of detectors. Consequently, the error bars on the \ac{WD} zero points include a systematic uncertainty contribution arising from residual detector-to-detector offsets.

To improve the consistency of flux measurements throughout the DR1 period, we adopted a step-wise approach, selecting a set of \ac{IS} values that track the throughput evolution derived from all sources in the self-calibration field, as shown in \cref{fig:WD_ZP_selfcal}.
The resulting zero-point residuals are shown in the right panel of \cref{fig:WD_ZP_selfcal}, together with additional measurements from the contamination monitoring programme that was executed on the centre of the self-cal field during early science operations \citep[F-014, see table D.1 in][]{DR1-TP026}. Since F-014 observations are based on a sequence of four dithers instead of 76, they are not suitable for deriving calibration products, but provide a useful independent data set for monitoring zero-point residuals. 
\Cref{fig:WD_ZP_selfcal} shows that the adopted calibration products provide a temporal stability of the \ac{NISP} photometric relative calibration that complies with the 1.5\% requirement. For most of the survey, the residuals remain at the level of approximately 2\,mmag. The main exception is the period between November 2023 and March 2024, during which no observations of the self-calibration field suitable for deriving calibration products were performed.
In DR2, a smooth interpolation model will be used to describe the throughput evolution at any time in the survey.

\subsection{Accuracy of the relative photometric calibration}

The same observations of the self-calibration field used to monitor the overall \ac{NISP} throughput are also exploited to assess the accuracy of the NIR relative photometric calibration. For each source, we extract individual photometric measurements from all exposures of all self-cal visits, and compute the median flux and the \ac{MAD}. These statistics are then analysed as a function of source magnitude to assess the photometric repeatability and identify any magnitude-dependent systematics.

\begin{figure}
    \centering
    \includegraphics[width=\linewidth]{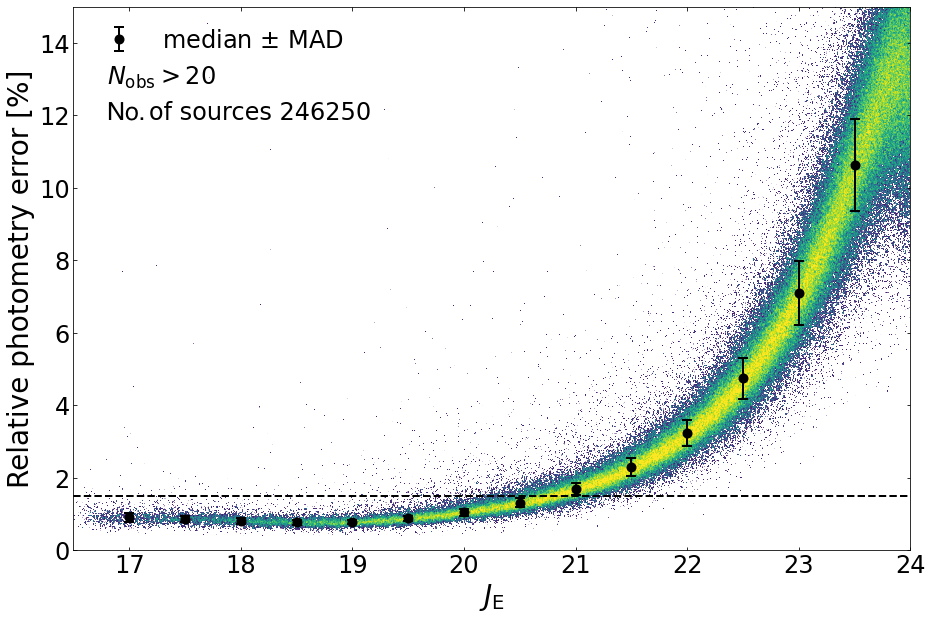}
    \caption{Relative flux calibration uncertainty for \JE estimated from all visits of the self-calibration field. The error for sources brighter than 19\,mag is at the level of 0.9\%. Similar results are obtained for \YE and \HE. The dashed line shows the 
    1.5\% requirement (maximum allowed post-calibration error).}
    \label{fig:relcal}
\end{figure}

The results are shown in \cref{fig:relcal} for the \JE\ band, and include statistical and systematic residual effects from ice contamination and unmasked persistence. At the bright end, where the measurements are not dominated by shot noise, the scatter of the repeated measurements corresponds to a relative flux calibration uncertainty of 0.9\%, compared to a requirement of 1.5\%. Similar results are obtained for the \YE\ and \HE\ bands. It should be noted that all self-calibration field visits are analysed together; therefore, the measured scatter also accounts for any potential time-dependent variations in the photometry.

\begin{figure*}
    \centering
    \begin{minipage}[t]{0.98\textwidth}
    \centering
    \includegraphics[width=5.8cm]{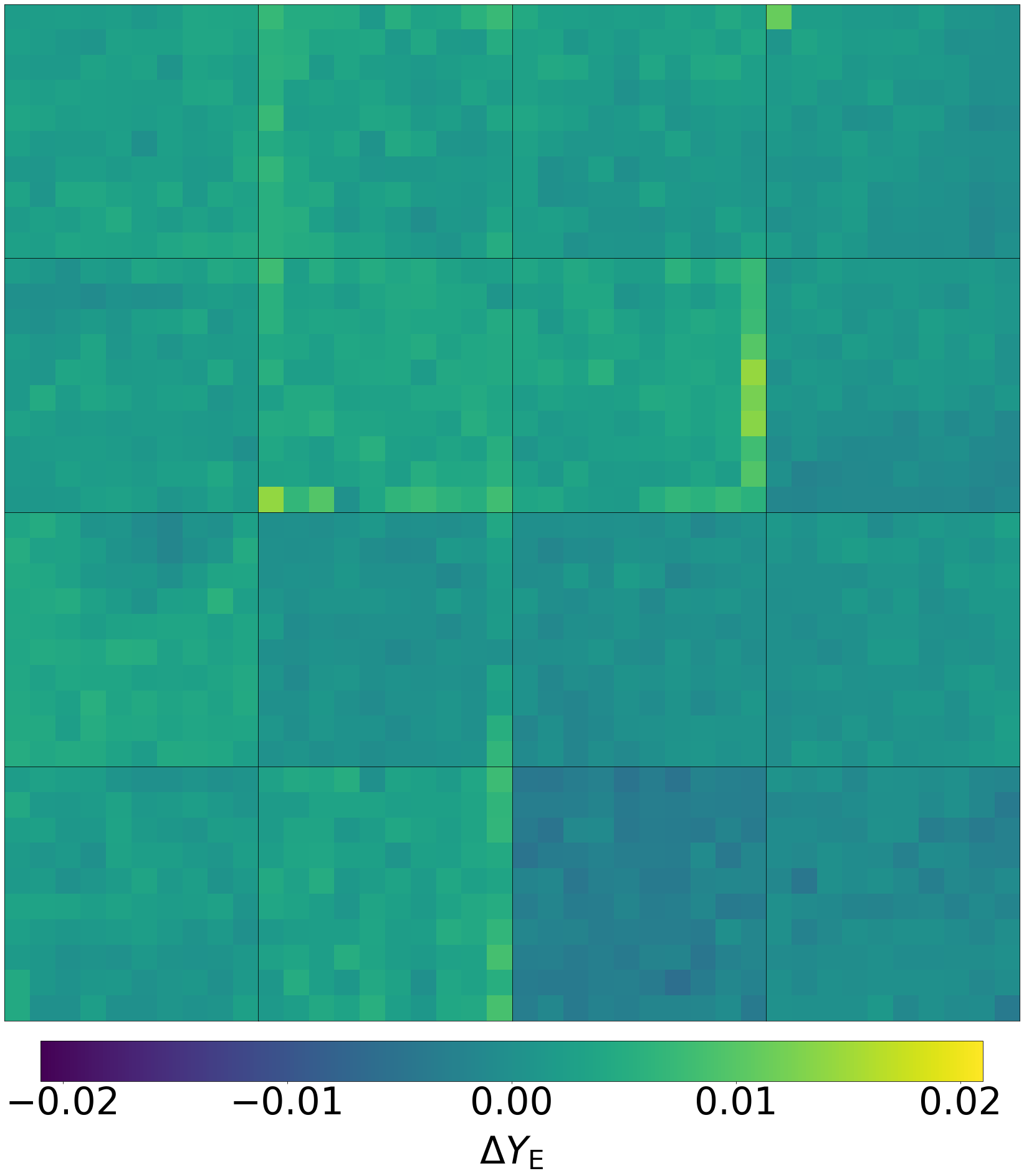}
    \includegraphics[width=5.8cm]{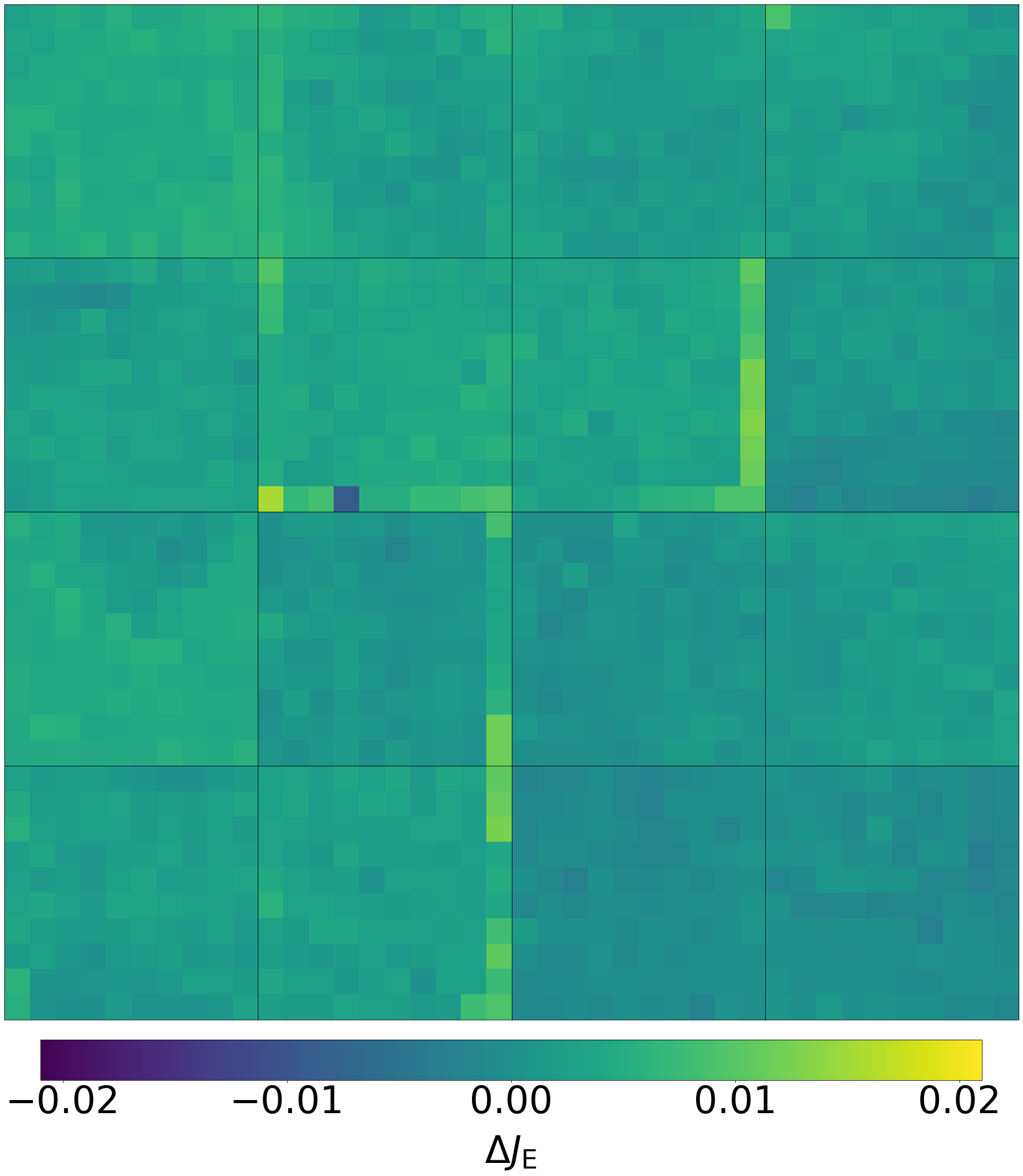}
    \includegraphics[width=5.8cm]{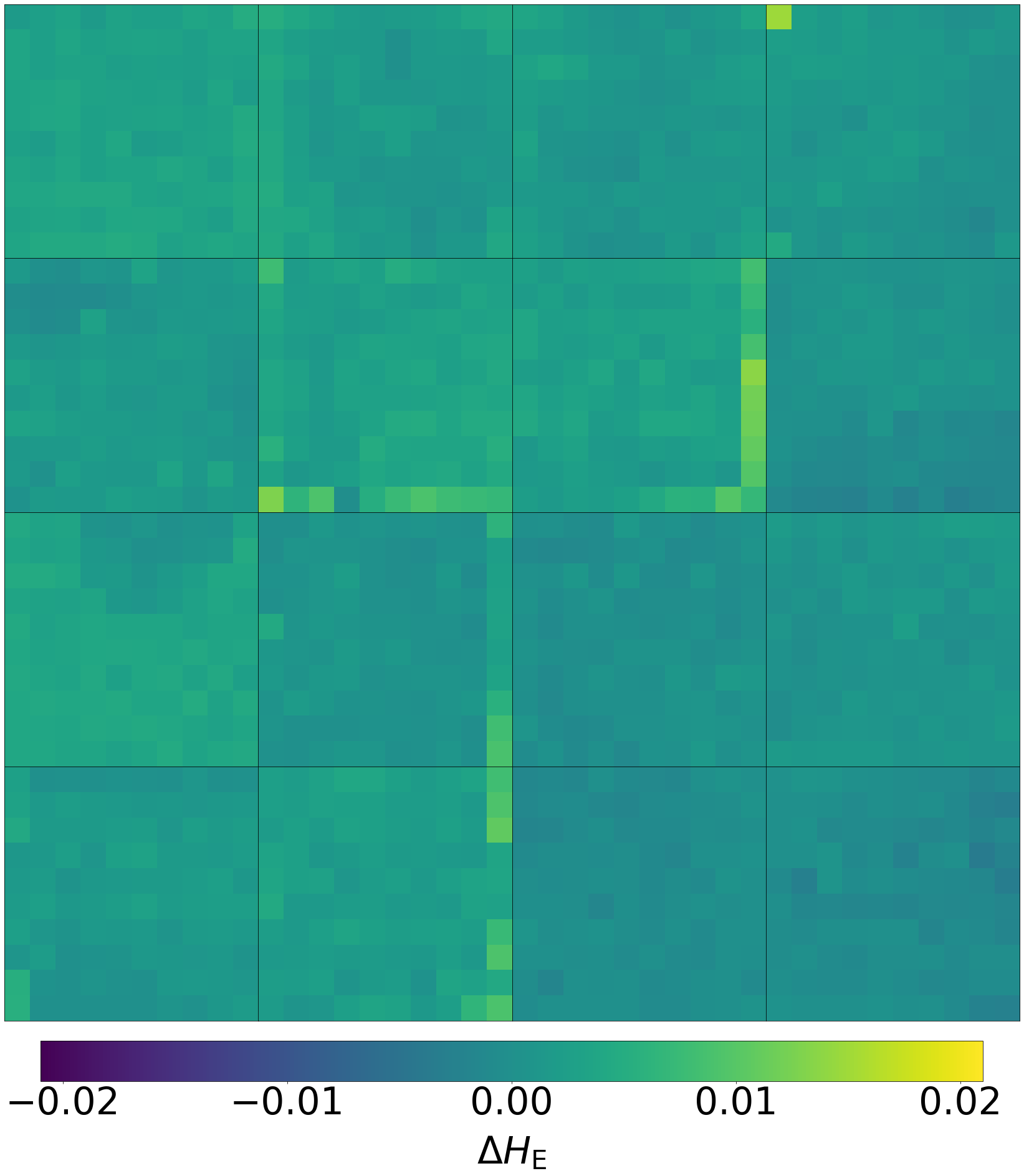}
    \end{minipage}
    \caption{\ac{FPA} distribution of photometric residuals for \YE, \JE, and \HE\ combining all observations of the self-calibration field during the DR1 period.} 
    \label{fig:FPA_residual}
\end{figure*}

The same data set can also be exploited to investigate potential spatial systematics by analysing photometric residuals as a function of detector position $(x,y)$ within the \ac{FPA}, using catalogue-level magnitude measurements from individual observations of the self-calibration field. The residuals are defined with respect to the mean magnitude of each source and projected onto the focal plane according to their detector coordinates.

Two-dimensional maps of the residuals are constructed by binning the measurements in detector coordinates and computing the median within each spatial bin. This approach mitigates the impact of outliers and enables the identification of coherent spatial structures across the detector.

The resulting residual maps are inspected for large-scale features that may reveal instrumental effects not fully corrected by the \ac{NIR PF} or the adopted calibration products. The analysis benefits from the high redundancy of the self-calibration field, where repeated measurements of the same sources sample different regions of the detector.
Results for the full data set, shown in \cref{fig:FPA_residual}, indicate that residual spatial features are at the level of 0.01\,mag or below.

\begin{figure}
    \centering
    \includegraphics[width=0.95\linewidth]{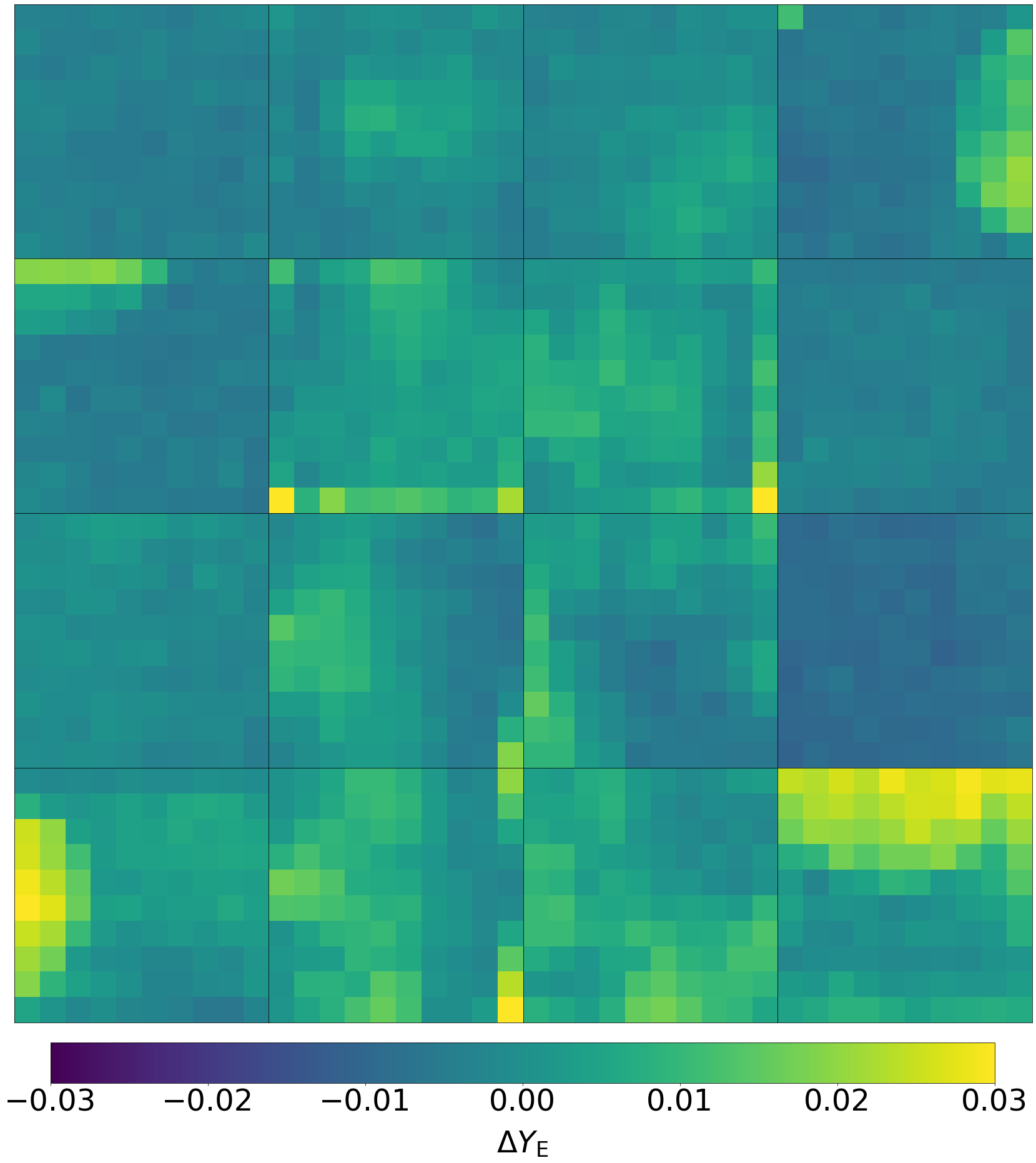}
    \caption{\ac{FPA} distribution of photometric residuals for \YE when considering only sources with 17 < \YE < 19. Similar results are obtained for \JE and \HE.}
    \label{fig:FPA_bright}
\end{figure}

When restricting the analysis to bright sources, however, the spatial distribution of the residuals across the \ac{FPA} reveals the features shown in \cref{fig:FPA_bright}, which closely resemble the persistence charge patterns shown in figure~4 of \citetalias{Q1-TP003}. This suggests that the flux-dependency of persistence parameters discussed in \cite{DR1-TP038}, and in particular the different behaviour in the bright and faint regimes, may provide an explanation for these features, as also noted in \citetalias{Q1-TP003}. Indeed, in terms of source counts, bright objects provide only a subdominant contribution to the large-scale flat-field estimation compared to the full source population.

\subsection{Accuracy of absolute photometric calibration}
\label{sec:abs_phot}
    
\begin{figure*}
    \centering
    \includegraphics[width=\textwidth]{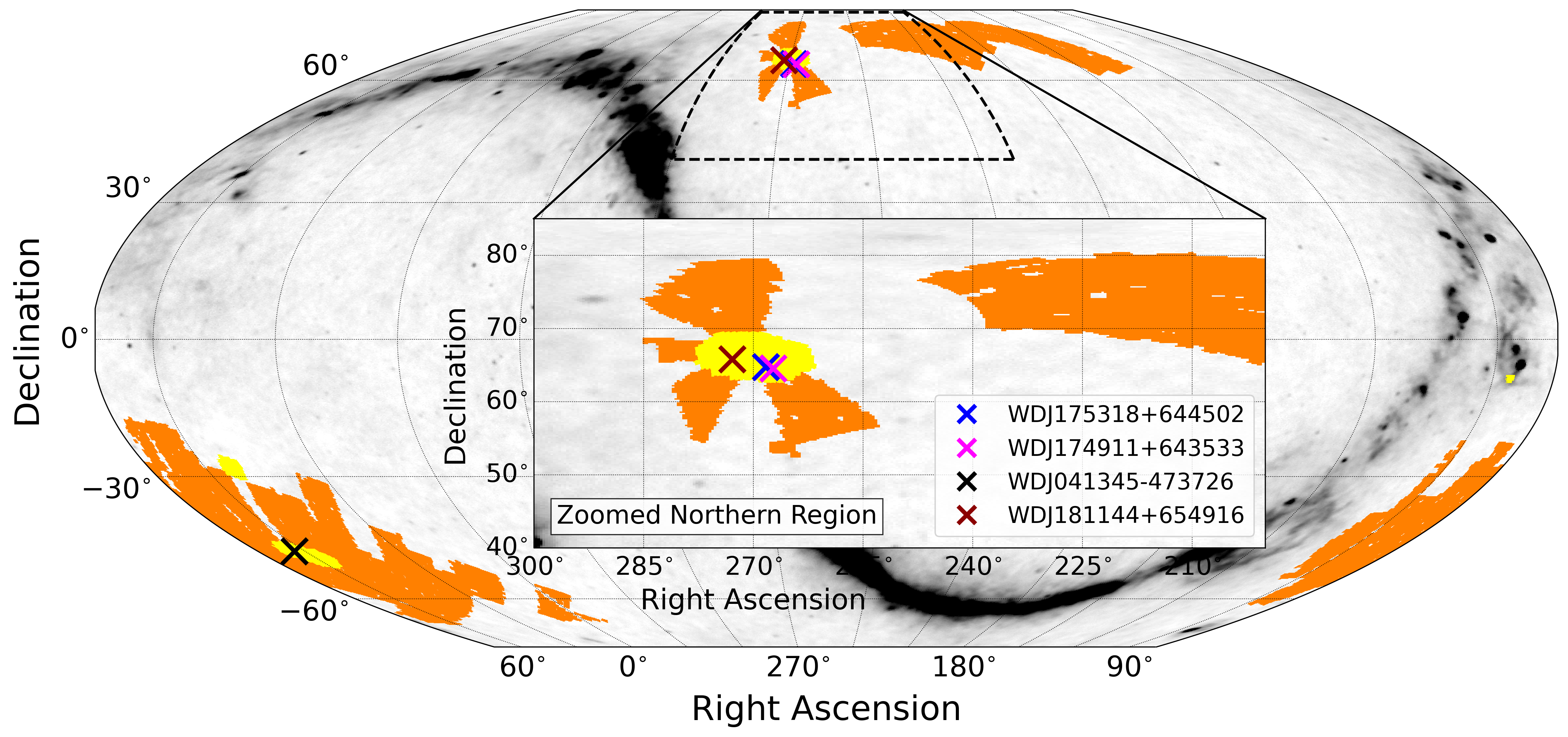}
    \caption{Full-sky map showing the positions of the four spectrophotometric \acp{WD} observed during the DR1 campaign. The main \ac{WD} calibrator used for the NISP absolute flux calibration is shown with a blue cross. The DR1 and Q1 coverages are shown in orange and yellow, respectively. }
    \label{fig:WD_skymap}
\end{figure*}

As discussed in \citetalias{Q1-TP003}, we adopt the faint spectrophotometric white dwarf WDJ175318.65+644502.15 ({\it Gaia} source ID 1440758225532172032), whose flux is tied to \ac{HST} CALSPEC \citep{bohlin2014,bohlin2017,bohlin2020}, as the primary calibrator for the flux calibration of the \Euclid\ NISP imaging. During the DR1 period, three other \Euclid spectro-photometric \acp{WD} were observed \citep[][; Deustua et al., in prep.]{DR1-TP026}, two in the EDF-North and one in EDF-South \citep{DR1-TP027,Scaramella-EP1}. \Cref{fig:WD_skymap} shows the positions of these three white dwarfs, together with that of WDJ175318.65+644502.15, relative to the footprints of the \ac{EWS} and \ac{EDS}, while Table~\ref{tab:wd_standards} summarises their coordinates and the NIR fluxes.

All three standards are tied to CALSPEC and have both photometric and spectroscopic HST observations obtained with WFC3/IR and STIS under programmes 16702 (PI: Appleton) and 17442 (Deustua). The hot \acp{WD} were originally selected from the {\it Gaia} DR2 \ac{WD} candidate catalogue \citep{GaiaWDcat} by requiring a temperature of $T > 11\,000$\,K, in order to exclude potentially variable cool \acp{WD} (Deustua et al., in prep.). Long- and short-term monitoring showed stability at the level of 1\% or better.  Observations of WDJ175318.65+644502.15 with VIS at optical wavelengths show that this particular standard in the self-cal field is probably more variable than initially thought; DR2 absolute flux calibration will be based on a sufficiently larger number of spectrophotometric standards \citep{DR1-TP026}.

\begin{table}
\centering
\caption{Spectrophotometric white dwarfs used for flux calibration and validation.}
\label{tab:wd_standards}
\setlength{\tabcolsep}{5pt}
\begin{tabular}{lccc}
\hline
\hline
\noalign{\vskip 1pt}
ID & RA & Dec  & $H_{\rm AB}$\\
\hline
\noalign{\vskip 1pt}
WDJ175318+644502 & 17:53:18.65 & +64:45:02.15 &  18.36 \\
WDJ174911+643533 & 17:49:11.78 & +64:35:33.54 & 18.26  \\
WDJ181144+654916 & 18:11:44.96 & +65:49:16.42  & 19.14  \\
WDJ041345$-$473726 & 04:13:45.06 & $-$47:37:26.29  & 18.27  \\
\hline
\end{tabular}
\end{table}

Following the analysis described in detail in \citetalias{Q1-TP003}, we derive the absolute flux calibration factor, or \ac{IS}, for these three additional white dwarfs and compare their values with that obtained for our primary calibrator, WDJ175318.65+644502.15. To assess the accuracy and temporal stability of the calibration, we examine the \ac{IS} values as a function of observation time, thereby accounting for potential temporal variations in filter throughput and photometric zero points.

\begin{figure}
    \centering
    \includegraphics[width=\linewidth]{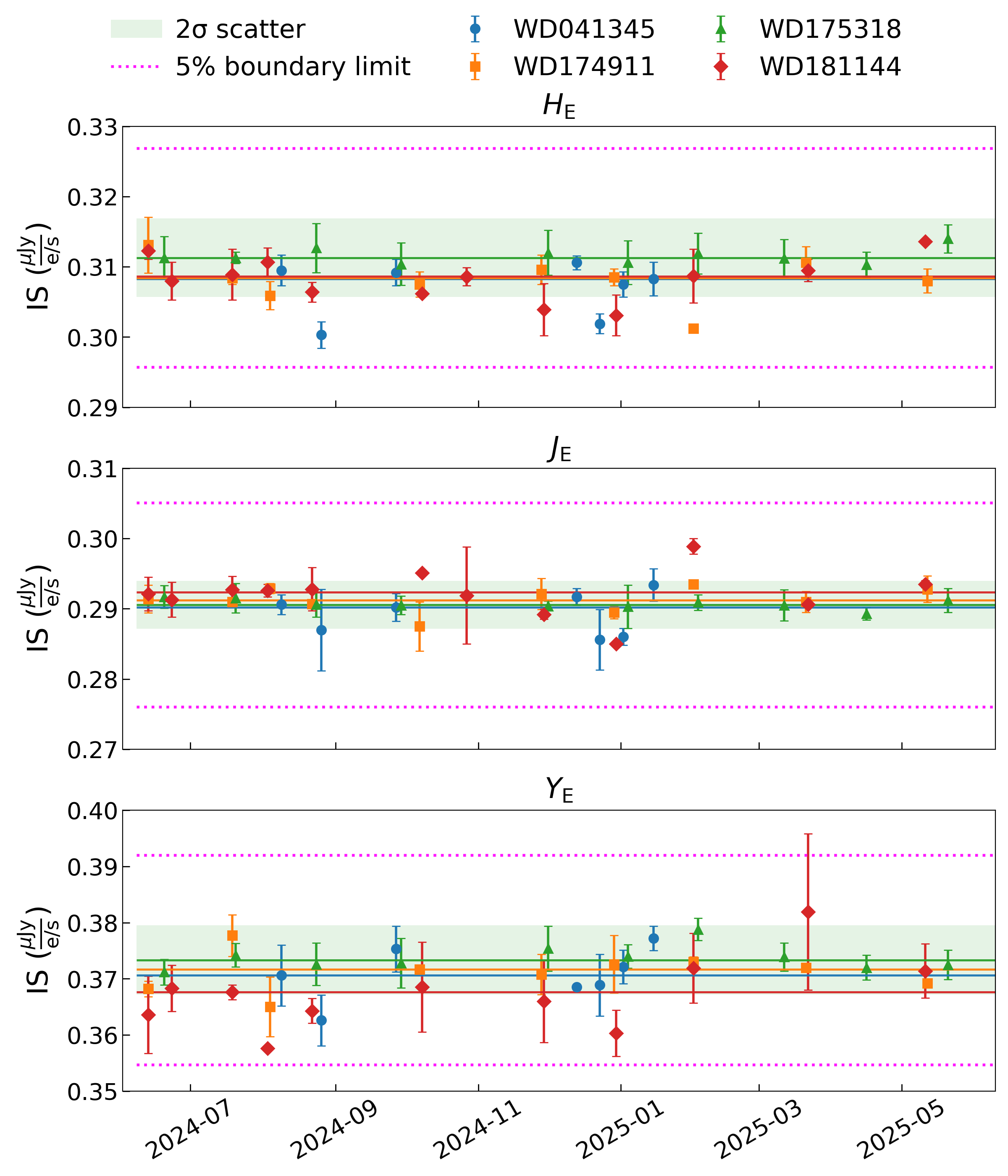}
    \caption{Inverse sensitivity as a function of observation time. The \ac{IS} measurements of the four hot \acp{WD}, including the main calibrator WDJ175318.65+644502.15, used for NISP absolute flux calibration are shown within the time range of June 2024 through June 2025. The median value of \ac{IS} measurements for the other three \acp{WD} are well within the $2\,\sigma$ scatter range of the main calibrator. We describe in the text how the scatter, $\sigma$, is measured using SMAD.}
    \label{fig:IS_allWDs_time}
\end{figure}

\Cref{fig:IS_allWDs_time} shows the \ac{IS} measurements for all four \acp{WD} in the three NISP bands. We restrict the time range to June 2024 through June 2025 to ensure sufficient temporal overlap among the comparison \acp{WD}. The primary calibrator, WDJ175318.65+644502.15, is shown in green. For each \ac{WD}, the median \ac{IS} value over the full time range is indicated by a solid line, with the line colour matching that of the corresponding data points. The shaded green region represents the $2\,\sigma$ scatter around the median IS of the primary calibrator. The scatter, $\sigma$, is estimated using the scaled median absolute deviation (SMAD), defined as $\mathrm{SMAD} = 1.4826 \times \mathrm{MAD}$, where MAD is the median absolute deviation.
 As shown in the figure, the median \ac{IS} values of the other three \acp{WD} lie well within the $2\,\sigma$ scatter of the primary calibrator. The dotted magenta line indicates the mission-required $5\%$ boundary.

\section{Discussion}
\label{sec:conclusions}

This paper has presented the \ac{NIR PF} used to process the \ac{NISP} raw photometric exposures for DR1. The analysis of the resulting calibrated single-epoch catalogues demonstrates that the main requirements for astrometric and photometric calibration are met. At the same time, a detailed validation of the images has highlighted a number of residual effects and limitations.
While these do not compromise the robustness of the scientific results obtained for the \Euclid DR1 core science programmes, they may be relevant for other scientific applications and analyses, such as low surface-brightness and galaxy-cluster studies. Their characterisation is therefore important for the proper interpretation of the data and their broader scientific exploitation, as well as to guide further improvements to the NIR data products in view of future releases, which will encompass larger datasets and require increasingly stringent levels of accuracy. 

\begin{itemize}
    \item Persistence. Although persistence modelling has improved significantly for DR1, it does not yet capture the full complexity of the effect \citep{DR1-TP038}, limiting the accuracy of procedures aimed at correcting the associated excess flux (Awad et al., in prep.). Therefore, a masking approach is preferred over attempting to correct the persistence signal. Tracing a 5-hour illumination history improves the ability to identify and mask residual streaks from spectroscopic observations. However, in some cases, this approach results in excessive masking. In particular: (i) in the presence of proton showers, energetic particle hits could remain masked for an extended period, causing the accumulation of masked regions and leading to the invalidation of several images; (ii) for DET13 and DET14, the masking results in the astrometric issues described in \cref{subsec:astrometry_res}. Although a solution has already been identified, it could not be implemented for DR1. VMPZ-ID {\tt InfoMaps} are therefore provided to identify the affected regions.
    
\smallskip
    
    \item Background estimation. As shown in \cref{fig:BkgSub}, while the outskirts of extended sources are preserved, residual horizontal stripes due to $1/f$ noise in the detector readout are still present. This is a known effect also observed in \ac{JWST}/NIRCam data \citep{JWST_NIRCam_1f} and will be addressed for DR2. We find that, when stacking a large number of background frames, the resulting master background shows spatial patterns similar to those seen in the persistence images and in the photometric residuals of bright sources shown in~\cref{fig:FPA_bright}. This indicates that the large-scale flat derived from source fluxes does not completely remove the corresponding background structure, consistent with the different levels of persistence contamination in the bright and faint regimes. When the dithering pattern is insufficient to average out these residual structures, they can result in an excess of diffuse faint signal. Finally, since the background is estimated independently for each detector, its uniformity on larger, focal-plane scales is not guaranteed.
    
\smallskip

    \item Relative and absolute photometric calibration. When considering the full sample of sources, the focal-plane distribution of photometric residuals shows no evidence of spatial features with a dispersion above the 0.01\,mag level. However, spatial structures become apparent when restricting the analysis to bright sources only. The absolute calibration is tied to a single-epoch observation of WDJ175318.65+644502.15, while the other secondary \acp{WD} observed by \Euclid\ are used for validation only and not used to improve the calibration accuracy. Temporal variations in the throughput due to the presence of ice are accounted for by adopting a set of \ac{IS} values in a step-wise approach. While the resulting residuals remain within the \Euclid\ requirements, this approach leaves room for further improvement. Moreover, the presence of ice affects the 16 NISP detectors differently. This is accounted for through the monthly estimation of detector offsets from self-calibration field observations, together with the large-scale flat. Inter-detector variations are not corrected, although the analysis presented in \cref{sec:masterflat} suggests that their impact is well below 1\% over most of the focal plane.
    
\smallskip

    \item Unmodelled instrumental eﬀects. These include spikes from bright stars and snowballs, which are partially masked during pre-processing, although no dedicated masking step is currently implemented. Arcs from bright stars \citep[see figure~21 in][]{EuclidSkyNISP} are not masked in DR1. Current investigations show the arcs are generally caused by bright stars within a particular area on the \ac{FPA}. From the DR1 data, preliminary work indicates that around 3\% of images show these arcs, with the exact position and shape dependent on the position of the bright star causing them. The exact optical surfaces causing these arcs and a model to describe them in order to provide masks, are under investigation and planned for future releases.

\end{itemize}

The analysis presented in this paper has identified a number of residual effects and limitations in the DR1 processing. Nevertheless, the results confirm that the DR1 NIR data products meet the required levels of astrometric and photometric accuracy and provide a robust basis for the scientific exploitation of the \Euclid\ data. The improvements being developed for DR2 are expected to further mitigate these effects and improve the quality and consistency of the NIR data products.

\begin{acknowledgements}
\AckEC
\AckDRone
This work has made use of data from the European Space Agency (ESA) mission
{\it Gaia} (\url{https://www.cosmos.esa.int/gaia}), processed by the {\it Gaia}
Data Processing and Analysis Consortium (DPAC,
\url{https://www.cosmos.esa.int/web/gaia/dpac/consortium}). Funding for the DPAC
has been provided by national institutions, in particular the institutions
participating in the {\it Gaia} Multilateral Agreement.
\end{acknowledgements}

\bibliographystyle{aa}
\bibliography{Euclid, main}

\begin{appendix}

\twocolumn[
    \section{Statistics for the non-transient BPM }
    
    \refstepcounter{table}
    \label{table:badpixdarks}      
    
    \noindent
    \textbf{Table~\thetable.}    NISP detector non-transient DQ flags in DR1.
    \begin{center}

    \centering          
    \begin{tabular}{c r r r r r r r r c c}     \hline\hline       
    \noalign{\vskip 1pt}

    DetID\tablefootmark{a} & \multicolumn{6}{c}{Invalid pixels\tablefootmark{b}} & \multicolumn{2}{c}{Informational\tablefootmark{b}} \\ 
    \hline
    \noalign{\vskip 1pt}
    & DISCONN & ZEROQE & SUPERQE & BADBASE & SAT & FLOWER & LOWQE & HOT \\ 
    \hline   
    \noalign{\vskip 1pt}                 
    11 & 450  & 747  & 290  & 27   & 15   & 3926  & 591 & 3601  \\  
    12 & 3655 & 5399 & 7038 & 25   & 1    & 15582 & 953 & 10756 \\  
    13 & 936  & 1261 & 551  & 74   & 54   & 6360  & 219 & 3661  \\  
    14 & 849  & 1858 & 579  & 183  & 155  & 22047 & 597 & 12830 \\  
    21 & 728  & 1275 & 16   & 171  & 115  & 7423  & 292 & 4714  \\  
    22 & 2971 & 3510 & 1313 & 2127 & 92   & 9881  & 331 & 8029  \\  
    23 & 1811 & 1992 & 2373 & 75   & 51   & 7105  & 605 & 3961  \\  
    24 & 490  & 698  & 302  & 15   & 10   & 4270  & 216 & 2160  \\  
    31 & 1721 & 2054 & 992  & 72   & 36   & 6870  & 725 & 7747  \\  
    32 & 3399 &  895 & 307  & 76   & 128  & 25274 & 420 & 5816  \\  
    33 & 768  & 1288 & 644  & 95   & 89   & 7837  & 402 & 7671  \\  
    34 & 1388 & 2072 & 249  & 105  & 11   & 6113  & 390 & 5764  \\  
    41 & 3220 & 1858 & 339  & 63   & 17   & 18604 & 584 & 2818  \\  
    42 & 663  & 847  & 404  & 62   & 4800 & 143   & 390 & 4086  \\  
    43 & 955  & 1129 & 401  & 17   & 3    & 2672  & 278 & 1997  \\  
    44 & 6016 & 7420 & 558 & 707   & 33   & 5588  & 1542& 11652 \\  

    \hline                  
    \end{tabular}
    \tablefoot{
    \tablefoottext{a}{Detector slot position in the NISP \ac{FPA}; see figure~A.1 in \citet{EuclidSkyNISP}.}
    \tablefoottext{b}{Pixel flag column names correspond to metadata keywords in the \ac{DQ} layer within each data product.}
    }
    \end{center}
]

    The statistics of the non-transient \ac{DQ} flags identified for the NISP detectors in DR1 are reported in \cref{table:badpixdarks}.

\section{Additional validation of the small-scale flat-field stability}
\label{app:prnu-stability}

\subsection{Masks, temporal sampling, and statistical checks}
\label{app:prnu-methods}

The monitoring sequence in \cref{sec:masterflat} contains one \HE master flat per month from March 2024 to April 2026, except for November 2024 and November 2025, for which no product is available. Temporal separations are calculated from the observing dates; the missing months are not interpolated.

For each detector, the common temporal mask retains pixels that are finite, positive, have zero data-quality (DQ) value, and lie outside the vignetting mask at every epoch. We dilate the excluded regions by 3 pixels before constructing the small-scale maps. The zero-DQ requirement excludes all flagged pixels, including informational flags, and is therefore stricter than excluding only the \texttt{INVALID} bit described in \cref{sec:badpix}. The retained fraction has a median of 96.25\% across detectors and a range of 90.62--98.81\%. These measurements describe the pixels retained at every epoch; they do not measure changes in the population of newly flagged pixels.

In the numerical calculation, we divide each master flat by its median over the common mask before smoothing. The mask-aware Gaussian operator smooths the valid-pixel-weighted image and the validity weights separately and divides the two results. Because the mask is fixed, the detector-median normalisation cancels in the ratio that defines $S_d$. Equation~\eqref{eq:prnu-small-scale} is therefore equivalent to the implemented calculation.

The reference-relative residual is $\Delta\ln S_{d,\mathrm{early}}(\mathbf{x},t)=\ln[S_d(\mathbf{x},t)/S_{d,\mathrm{early}}(\mathbf{x})]$. The first three epochs enter $S_{d,\mathrm{early}}$ and are not independent of it. Their residuals are retained in the diagnostic heatmap but omitted from the reference-relative time-series plot. The adjacent-epoch statistic uses consecutive available products, including the longer intervals across the missing months. The late-to-early comparison uses two disjoint three-epoch median stacks.

The structure function uses all 276 distinct epoch pairs, with 42, 53, 90, 65, and 26 pairs in the lag bins 0--90, 90--180, 180--365, 365--550, and 550--800 days, respectively. Each pair contributes one measurement per detector. Pairs sharing an epoch are not independent. For the jackknife calculation, we omit each epoch in turn, remove every pair containing that epoch, and recalculate the lag-bin medians. The resulting jackknife standard errors describe sensitivity to the sampled epochs, not the unknown contributions of the calibration source or processing.

\begin{figure*}
\centering
\includegraphics[width=1\textwidth]{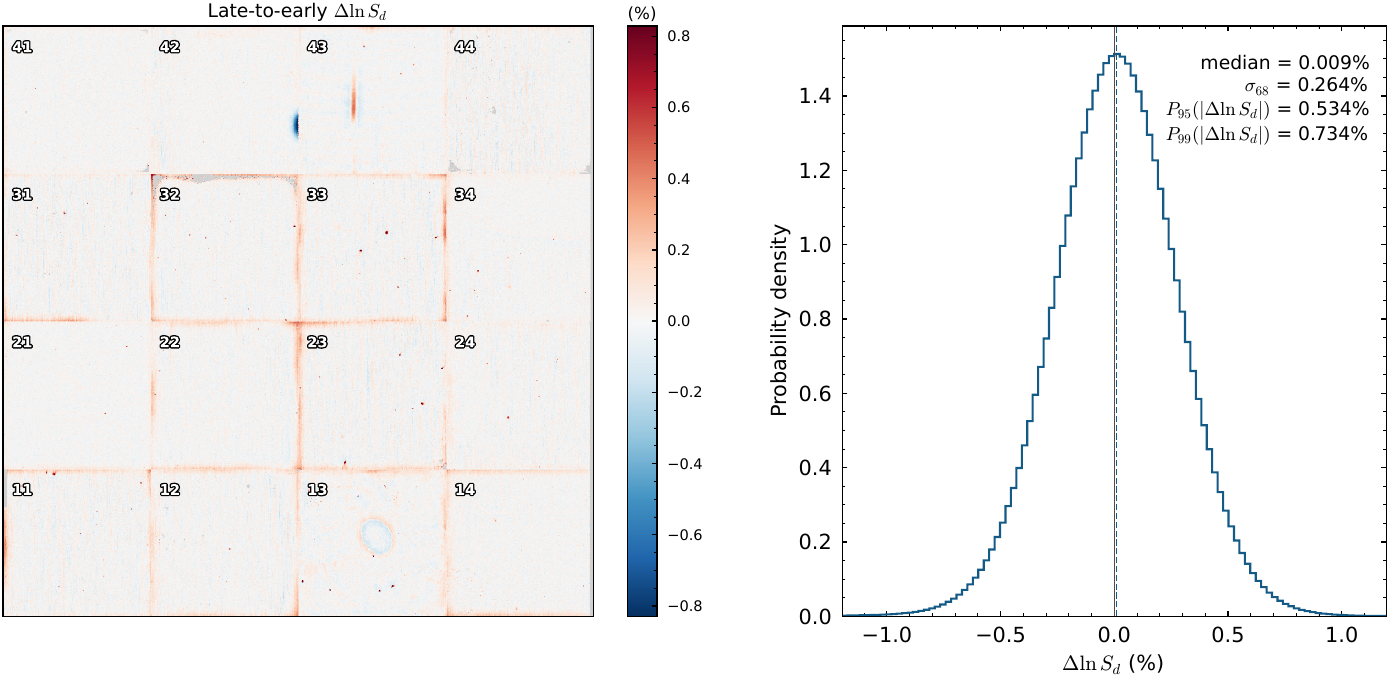}
\caption{Late-to-early small-scale logarithmic residual, comparing pixel-wise median stacks of the final three and first three epochs. \emph{Left:} The focal-plane mosaic, binned 8\,$\times$\,8 for display only. The colour scale is symmetric about zero, detector residual medians are retained, and pixels excluded by the common temporal mask are shown in neutral grey. No additional detector-edge exclusion is applied. \emph{Right:} distribution and annotated statistics of all unbinned valid focal-plane pixels.}
\label{fig:flat-stability-fpa-v2}
\end{figure*}

Detector medians and pooled focal-plane statistics are calculated separately. For example, the range 0.246--0.291\% describes the 16 detector-level late-to-early scatters, whereas 0.264\% is calculated from the pooled pixel distribution. The latter distribution also supplies the percentile annotations in \cref{fig:flat-stability-fpa-v2}; its median residual is 0.009\%.

The main residual mosaic and its statistics use the same common mask, with no additional edge exclusion. As a separate check, excluding the outer 64 pixels of each detector changes the pooled late-to-early $\sigma_{68}$ from 0.2640\% to 0.2620\%, a relative difference of 0.76\%. Thus, these edge regions do not dominate the pooled scatter. Changing the Gaussian standard deviation for the smoothing from 64 to 32 or 128 pixels changes the final reference-relative scatter by less than 1\% relative to the nominal result. The morphology of narrow boundary haloes remains scale dependent, so the stability of the global statistic does not establish the origin of individual boundary structures.

\subsection{Measurements of residual structures}
\label{app:prnu-features}

All measurements in this section use the late-to-early logarithmic residual on the native detector grid, before the display rotations and 8\,$\times$\,8 binning in \cref{fig:flat-stability-fpa-v2}. Positive and negative components are identified separately. A local contrast is the median residual within a selected region minus the median residual in its surrounding control region. These contrasts describe structures in the delivered master flats and are not assigned a physical origin.

\paragraph{Detector boundaries and extended features.}
For each of the four edges of each detector, we form a profile of the median residual as a function of distance from the edge. We compare the median of this profile over distances 0--7 pixels with its median over an interior control interval of 32--63 pixels, with both intervals chosen according to the adopted binning and Gaussian smoothing scales. A positive edge excess is selected when the difference exceeds both 0.05\% and three times the $\sigma_{68}$ of the control-profile values. This criterion selects 27 of the 64 edges, distributed over ten detectors. Among these selected edges, the median edge-minus-control contrast is 0.087\%, with 16th--84th percentiles of 0.066--0.114\%. These are empirical profile-selection thresholds, not false-alarm probabilities.

The negative feature on detector 42 intersects a detector edge. Its selected footprint spans 74\,$\times$\,291 native pixels and has a median local contrast of $-0.499\%$. The dimensions describe the observed, edge-truncated footprint, not the full extent of the feature. Detector 43 contains a vertical positive band, and detector 13 contains a broad negative ellipse with a positive surround (see \cref{fig:flat-stability-fpa-v2}). Given the amplitude of the detected features and the small fraction of pixels involved across the three regions, their impact on the overall quality of the NIR images is negligible. In terms of their local impact, the measured contrasts are well within the 1.5\% relative calibration accuracy requirement.

\paragraph{Compact positive regions.}
We construct a detection map by subtracting the late-to-early residual smoothed with a Gaussian of $\sigma=12$ pixels from the same residual smoothed with $\sigma=1.5$ pixels. Both convolutions are normalised by their valid-pixel support. For each detector, we estimate the median and a robust scale from control pixels at least 24 pixels from detector edges and 8 pixels from invalid regions. The scale is the median absolute deviation multiplied by 1.4826, with iterative clipping at 4.5 times that scale. This detection scale differs from the $\sigma_{68}$ used for the stability measurements. It includes correlated residual structure and is not a measurement of master-flat noise.

We identify connected positive regions above the control median by at least 2.5 times the detection scale, retaining regions that contain a pixel at least 5 times that scale above the median. Connectivity includes the eight neighbouring pixels. The compact-region selection requires at least nine valid pixels, a second-moment ellipse axis ratio no greater than 3, and a minor-axis length of at least 1 pixel. Regions with a major-axis or bounding-box extent of 80 pixels or more are classified separately as elongated structures. We exclude footprints that touch a detector edge or invalid pixel, overlap the targeted positive surround on detector 13, or lie within 2 pixels of an opposite-sign component with an unsmoothed negative residual below $-0.25\%$. These feature-selection restrictions do not alter the common mask used for the global stability measurements or the main mosaic.

For a selected footprint with area $A$ in native-pixel units, we define the area-equivalent diameter as
\begin{equation}
D_{\mathrm{eq}}=2\sqrt{A/\pi}\,.
\label{eq:prnu-equivalent-diameter}
\end{equation}
We measure the local contrast on the original, unsmoothed residual using the selected footprint and a surrounding control annulus formed by binary dilations of 5 and 18 pixels. Invalid pixels and other detected components are excluded from the control annulus. Detection smoothing therefore determines the footprint, but not the residual values used to measure its contrast.

The nominal selection contains 639 compact positive regions across all 16 detectors. Their area-equivalent diameters have a median of 5.4 pixels, with 16th--84th percentiles of 4.2--7.6 pixels. Their local contrasts have a median of 0.22\%, with 16th--84th percentiles of 0.09--0.33\%. These intervals describe the selected population, not uncertainties on its medians. The diameters describe thresholded footprints and depend on the detection smoothing and selection thresholds; they are not intrinsic detector-defect sizes. Most selected regions are smaller than the 8-pixel display bins in \cref{fig:flat-stability-fpa-v2}; their sizes are measured from the native arrays, not from that mosaic.

\subsection{Detector-wide changes in master-flat normalisation}
\label{app:prnu-detector-scale}

\begin{figure*}
\centering
\includegraphics[width=1\textwidth]{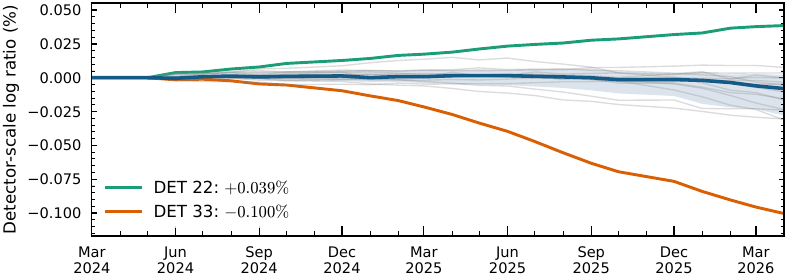}
\caption{Detector-wide logarithmic changes in the master flats before median normalisation. Thin grey curves show individual detectors, while the blue curve and shaded band show the median and 16th--84th percentiles across detectors. Detectors 33 and 22 have the largest negative and positive changes at the final epoch and are highlighted. The ordinate is $100\,m_d(t)$ in percent, with $m_d(t)$ defined in \cref{eq:prnu-detector-scale}. A detector-wide multiplicative change cancels from the small-scale metric.}
\label{fig:detector-scale-v2}
\end{figure*}

We measure detector-wide changes separately from the small-scale residuals. Let $\mathcal{F}_{d,\mathrm{early}}(\mathbf{x})$ be the pixel-wise median of the first three master flats before detector-median normalisation. We define
\begin{equation}
m_d(t)=\operatorname{median}_{\mathbf{x}\in\Omega_d}
\ln\!\left[
\frac{\mathcal{F}_d(\mathbf{x},t)}
{\mathcal{F}_{d,\mathrm{early}}(\mathbf{x})}
\right]\,.
\label{eq:prnu-detector-scale}
\end{equation}
At the final epoch, the median of $100\,m_d(t)$ across detectors is $-0.008\%$, with a range from $-0.100\%$ on detector 33 to $+0.039\%$ on detector 22 (\cref{fig:detector-scale-v2}). These quantities describe detector-wide scale changes in the delivered master flats, not an isolated measurement of detector sensitivity. A purely multiplicative detector-wide change cancels in \cref{eq:prnu-small-scale} and is therefore absent from the small-scale metric.

\begin{figure*}
\centering
\includegraphics[width=1\textwidth]{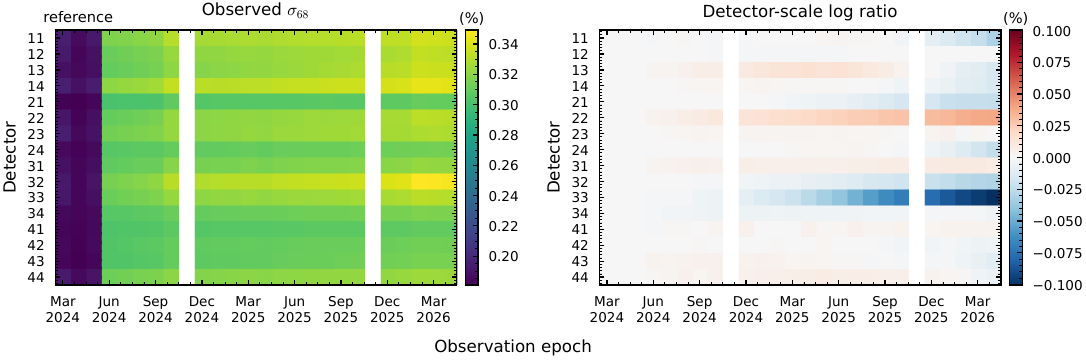}
\caption{Detector-by-epoch diagnostics, with epochs in chronological order and detectors ordered by identifier. \emph{Left:} Reference-relative scatter $\sigma_{68}(\Delta\ln S_{d,\mathrm{early}})$. The dashed line marks the end of the first-three-epoch reference interval. Residuals within that interval share data with the reference and are not comparable to those of subsequent epochs. \emph{Right:} Detector-wide logarithmic change, $100\,m_d(t)$. Both panels are in percentages and have separate colour scales. Blank columns mark November 2024 and November 2025, for which no product is available.}
\label{fig:detector-epoch-v2}
\end{figure*}

Figure~\ref{fig:detector-epoch-v2} shows both the reference-relative small-scale scatter and the detector-wide scale change for each detector and epoch. The low reference-relative scatters in the first three columns result from including those epochs in the reference stack. They must not be interpreted as evidence for a transition at the fourth epoch. The blank columns mark the two months without a master-flat product.

\section{Description of NIR DPs}
\label{app:dataproducts}

All \acp{DP}, defined in the Euclid Common Data Model as XML files, include two main metadata elements: {\tt DataStorage}, which provides information on the associated \ac{MEF} file, and {\tt QualityParams}, which contains statistics extracted from the \ac{MEF} file to assess the quality of the \ac{DP} and validate its production.

\subsection{Calibrated single-epoch exposures}

The {\tt DpdNirCalibratedFrame} provides fully astrometrically and photometrically calibrated individual exposures; its structure is described in \citetalias{Q1-TP003}. In DR1, the \ac{DP} structure has been extended with the new {\tt PsfSingleStorage} element, in addition to the {\tt DataStorage}, {\tt BackgroundStorage}, and {\tt PsfStorage} elements already present in Q1:
\begin{itemize}
    \item {\tt DataStorage} -- references the MEF file containing the calibrated image data;
    \item {\tt BackgroundStorage} -- references the MEF file containing the background measured as described in \cref{sec:background};
    \item {\tt PsfStorage} -- references two FITS files, one containing the \ac{PSF} model and one containing the \ac{PSF} image, obtained from observations of the self-calibration field as described in \cref{sec:psf} and \citet{DR1-TP005};
    \item {\tt PsfSingleStorage} -- references the \ac{PSF} model and image derived from the calibrated image itself. 
\end{itemize}

\subsection{Calibrated single-epoch catalogues}

The {\tt DpdNirCalibratedFrameCatalog} is the source catalogue extracted from the calibrated single-epoch exposure (\cref{sec:catalog}). Its global metadata structure therefore mirrors that of the {\tt DpdNirCalibratedFrame}.

The \ac{MEF} file is structured into 17 \acp{HDU}: a primary \ac{HDU} and 16 detector-specific HDUs, each containing a catalogue table with a wide range of parameters for the detected sources, including fluxes and magnitudes estimated from multi-aperture photometry, PSF fitting, and {\tt SourceExtractor MAG\_AUTO}. It is important to note that aperture corrections are not applied at this stage. While magnitudes are provided in the AB system, fluxes are still reported in units of $\mathrm{e^-}$.

\subsection{Stacked exposures}

The {\tt DpdNirStackedFrame} is the result of the resampling and stacking process described in \cref{sec:resandstack}, combining 4 (76) single-epoch images for \ac{EWS} (self-calibration field) observations. Its MEF file consists of a single {\tt SCI} \ac{HDU} covering the entire stacked science image, together with the corresponding {\tt RMS} and Data Quality {\tt DQ} extensions. Further details on the resampling and stacking process are provided in \cref{sec:resandstack}.

The \ac{DP} metadata elements collect all information describing the stacked observation, including the {\tt Filter}, the {\tt WCS} keywords used to locate the image on the sky, the {\tt ZeroPoint} containing the effective photometric image zero point, and the {\tt DateRange} indicating the time interval covered by the observations contributing to the stacked image.

The MEF files are referenced by the same {\tt DataStorage}, {\tt PsfStorage}, and {\tt PsfSingleStorage} elements used for the single-epoch images. The {\tt BackgroundStorage} element is not present, since the background is subtracted from the images during the stacking process (see~\cref{sec:resandstack}). All photometric calibration factors are applied during the process, and therefore the stacked image is in units of $\mathrm{\mu Jy}$.

As for the calibrated single-epoch frame, the {\tt QualityParams} metadata element contains a set of statistics for the three MEF files described above. In this case, however, the internal structure is simpler than that of the single-epoch image, since it contains a single layer corresponding to the entire focal plane. 

\subsection{Catalogues from NIR stacks}

The catalogue extracted from the stacked image is stored in the {\tt DpdNirStackedFrameCatalog} DP. Its metadata structure is similar to that of the stacked image, with the {\tt DataStorage} element referencing a MEF file structured into two \acp{HDU}: a primary \ac{HDU} and a second \ac{HDU} containing a catalogue table with the parameters of all detected sources in the stacked image.

As for the single-epoch catalogue, aperture correction is not applied to either fluxes or magnitudes. Magnitudes are provided in the AB system, while fluxes are reported in $\mathrm{\mu Jy}$.

\end{appendix}

\end{document}